%% file: main.tex
\PassOptionsToPackage{unicode, pdfprintscaling=None, colorlinks, allcolors=blue!50!black}{hyperref}
\documentclass[prd,  floatfix, preprintnmbers,  unsortedaddress, superscriptaddress, longbibliography, nofootinbib]{revtex4-2}

\input{preamble.tex}
\input{acronyms.tex}

\begin{document}

\title{Vacuum Entanglement Harvesting in the Ising Model}
\author{Hersh Singh\,\orcidlink{0000-0002-2002-6959}}
\email{hershsg@uw.edu}
\affiliation{InQubator for Quantum Simulation (IQuS), Department of Physics, University of Washington, Seattle, WA} 
\author{Tanmoy Bhattacharya\,\orcidlink{0000-0002-1060-652X}}
\email{tanmoy@lanl.gov}
\affiliation{T-2, Los Alamos National Laboratory, Los Alamos, NM}
\author{Shailesh Chandrasekharan\,\orcidlink{0000-0002-3711-4998}}
\email{sch27@duke.edu}
\affiliation{Department of Physics, Duke University, Durham, NC}
\author{Rajan Gupta\,\orcidlink{0000-0003-1784-3058}}
\email{rajan@lanl.gov}
\affiliation{T-2, Los Alamos National Laboratory, Los Alamos, NM}

\preprint{LA-UR-21-26805, IQuS@UW-21-045, INT-PUB-23-004}

\begin{abstract}
The low-energy states of quantum many body systems, such as spin chains, are entangled. Using tensor network computations, we demonstrate a protocol that distills Bell pairs out of the ground state of the prototypical transverse-field Ising model.  We explore the behavior of rate of entanglement distillation in various phases, and possible optimizations of the protocol.  Finally, we comment on the protocol as we approach quantum criticality defining a continuum field theory.
\end{abstract}

\maketitle

\input{sec_intro}
\input{sec_negativity}
\input{sec_groundstate}
\input{sec_protocol_single}
\input{sec_protocol_repeat}
\input{sec_conclusions}

\bibliographystyle{apsrev4-2-errat}
\showtitleinbib
\bibliography{refs}

\appendix
\input{app_effdens}
\input{app_perm}
\input{app_mps}
\input{app_exact}
\input{app_alignment}

\printglossary

\end{document}

%% file: preamble.tex
\ifx\pdfsuppresswarningpagegroup\undefined\else\pdfsuppresswarningpagegroup=1\fi\relax

\usepackage{orcidlink}
\usepackage{multirow}
\usepackage{siunitx}

\usepackage{tikz}
\usetikzlibrary{decorations.markings, arrows, positioning, calc}

\usepackage{graphicx,times}
\usepackage{latexsym}
\usepackage{mathtools}

\usepackage{amsmath,amssymb,amsbsy,amsfonts}
\usepackage{array}
\usepackage{bm}
\usepackage{graphics}
\usepackage{mathrsfs}
\usepackage{xcolor}
\usepackage{cancel}
\usepackage[normalem]{ulem}

\usepackage{inconsolata}

\usepackage[capitalise, nameinlink]{cleveref}

\usepackage{xcolor}
\usepackage{makecell}
\newcommand\TopRule{\Xhline{0.08em}}

\newcommand\MidRule{\Xhline{0.03em}}
\newcommand\BotRule{\Xhline{0.08em}}

\newcommand{\sA}{s^A}
\newcommand{\sB}{s^B}
\newcommand{\sgA}{\sigma^A}
\newcommand{\sgB}{\sigma^B}
\newcommand{\sgC}{\sigma^C}

\renewcommand\>{\rangle}
\newcommand\<{\langle}

\DeclareMathOperator\Tr{Tr}

\usepackage{microtype}

\usepackage{enumitem}
\setlist[itemize]{leftmargin=4.0mm}

\usepackage{FiraSans}
\usepackage{FiraMono}

\usepackage{silence}
\usepackage{ctable}

\newcommand\HSp{\mathcal{H}}

\newcommand\ket[1]{\mathclose{\,\mathopen{|}\mathord{#1}\mathclose{\rangle}}}
\newcommand\bra[1]{\mathopen{\mathopen{\langle}\mathord{#1}\mathclose{|}\,}}

\newcommand\Id{\mathbb{I}}

\newcommand\rhoth{\ensuremath{\rho_{\mathrm{th}}}}
\newcommand\rhog{\ensuremath{\rho_{+}}}
\newcommand\ZZ{\mathbb{Z}}

\newcommand\GG{{\mathsf{G}}}
\newcommand\GGt{ {\tilde{\mathsf{G}}} }

\newcommand\Neg{\mathcal{N}}
\newcommand\deltaMax{\delta_{\text{max}}}

\newcommand\footnotepunct[2]{{\edef\tmp{\spacefactor=\the\spacefactor\relax#2}%
    \toks0={#1}%
    \hbox to 0pt{\tmp
      \edef\tmp{\noexpand\footnote{\the\toks0}%
        \spacefactor=\the\spacefactor\relax}%
      \hss\expandafter}%
    \tmp}}

\makeatletter
\newcommand\showtitleinbib{{\escapechar=`\\ \immediate\write\@auxout{%
\csname citation{REVTEX42Control}\endcsname^^J%
\csname citation{apsrev42Control}\endcsname
}}}
\makeatother



%% file: acronyms.tex
\usepackage{relsize}

\usepackage[acronyms, nohypertypes={acronym}, nopostdot, style=super, nonumberlist, toc]{glossaries}
\setacronymstyle{long-sc-short}
\glsenableentrycount
\makeglossaries  

\newcommand\ac[1]{\gls{#1}}

\newcommand\acp[1]{\glspl{#1}}

\newacronym{WF}{wf}{Wilson-Fisher}
\newacronym{AF}{af}{asymptotically free}

\newacronym{RG}{rg}{renormalization group}

\newacronym{QIS}{qis}{Quantum Information Science}
\newacronym{PPT}{ppt}{positive-semidefinite partial transpose}
\newacronym{NPT}{npt}{negative partial transpose}

\newacronym[longplural={conformal field theories}]{CFT}{cft}{conformal field theory}
\newacronym[longplural={lattice field theories}]{LFT}{lft}{lattice field theory}
\newacronym[longplural={effective field theories}]{EFT}{eft}{effective field theory}
\newacronym[longplural={quantum field theories}]{QFT}{qft}{quantum field theory}

\newacronym[]{DMRG}{dmrg}{Density Matrix Renormalization Group}
\newacronym[]{TFIM}{tfim}{Transverse Field Ising Model}

\newacronym[]{LOCC}{locc}{Local Operations and Classical Communicaton}
\newacronym[]{OBC}{obc}{open boundary conditions}

\newacronym{MPS}{mps}{matrix product states}

\newacronym{JLP}{jlp}{Jordan-Lee-Preskill}

\newacronym{BBN}{bbn}{big bang nucleosynthesis}

\newacronym{LEC}{lec}{low-energy constant}

\newacronym{QCD}{qcd}{quantum chromodynamics}
\newacronym{MC}{mc}{Monte Carlo}

\newacronym{IR}{ir}{infrared}
\newacronym{UV}{uv}{ultraviolet}

\newacronym{QED}{qed}{quantum electrodynamics}
\newacronym{SNR}{snr}{signal-to-noise ratio}

\newacronym{NLSM}{nlsm}{nonlinear sigma model}

\newacronym{CL}{cl}{Complex Langevin}

\newacronym{CSA}{csa}{Cartan subalgebra}

\newacronym{SSB}{ssb}{spontaneous symmetry breaking}

\newacronym{AFQMC}{afqmc}{auxiliary field quantum Monte Carlo}
\newacronym{iHMC}{ihmc}{imaginary-mass Hybrid Monte Carlo}

\newacronym{MCMC}{mcmc}{Markov Chain Monte Carlo}

\newacronym{QI}{qi}{quantum information}

%% file: sec_intro.tex
\section{Introduction}
\label{sec:intro}

Continuum \acp{QFT} are the cornerstone for our current understanding of the standard model of particle physics. Motivated by experiments, traditional studies of \acp{QFT} had focused on calculating the phases of ground states through local order parameters, spectral quantities of low lying excitations, interaction coupling constants, etc. More recently, there has been considerable interest in understanding the connection of \acp{QFT} to quantum information science.
This has motivated physicists to consider entanglement properties of \ac{QFT} \cite{RevModPhys.90.045003}. While it has been known for a long time that bounded energy states of \acp{QFT} display long-distance entanglement~\cite{Reeh1961,Schlieder1965,PhysRevD.14.870,SUMMERS1985257,doi:10.1063/1.527733,doi:10.1063/1.527734,Redhead1995,Redhead1998,PhysRevA.58.135}, recently the focus has turned to understanding the entanglement in interacting quantum many body systems~\cite{Cardy08,*[{See review }] [{ and references therein.}] RevModPhys.80.517}. The nature~\cite{PhysRevLett.96.110404,PhysRevLett.96.110405} and the dynamics~\cite{Calabrese_2005,PhysRevB.95.094302} of this entanglement has been extensively studied in a limited class of systems.
Entanglement properties of ground states can also help identify new types of phases of matter, based on topological order that are difficult to understand using local order parameters \cite{Wen:1989iv}.
Entanglement has also been shown to be key to understanding the success of tensor network algorithms for quantum many body systems \cite{cirac_matrix_2020}.

While entanglement structure has been well established to be important in gaining insight into quantum systems with extensive degrees of freedom, such as \acp{QFT}, one can ask whether this entanglement can be used to carry out useful tasks?  Many quantum information protocols, such as communication \cite{nielsen00} and  teleportation \cite{bennett_teleporting_1993}, rely on a supply of entangled Bell pairs.
The idea of extracting Bell pairs from the vacuum of a \ac{QFT} was pioneered in Refs.~\cite{valentini_1991, PhysRevA.71.042104, Reznik2003, Reznik:2000mm}. In particular, these early works demonstrated that two qubits can be entangled by coupling them to quantum fields, even in a spacelike separated region of spacetime. Further works have analyzed 
entanglement harvesting from thermal states~\cite{brown_thermal_2013, braun_entanglement_2005, braun_creation_2002}, from coherent scalar field states~\cite{simidzija_nonperturbative_2017}, and from the electromagnetic vacuum using hydrogen-like atoms~\cite{pozas-kerstjens_entanglement_2016}; and investigated
dependence of entanglement harvesting on switching protocols~\cite{pozas-kerstjens_harvesting_2015a}, on the detector gaps in relation to the mass of the field~\cite{maeso-garcia_entanglement_2022}, on other detector properties~\cite{pozas-kerstjens_harvesting_2015a, salton_accelerationassisted_2015, sachs_entanglement_2017} and on the spacetime structure~\cite{steeg_entangling_2009, martin-martinez_spacetime_2016}.

While these studies have demonstrated the entanglement in physical states of \acp{QFT} can indeed be exploited for quantum information tasks~\cite{valentini_1991,Reznik:2000mm,Reznik2003,PhysRevA.71.042104}, they are formulated in the continuum and most limit themselves to perturbation theory.
For a fundamental understanding of the phenomena and for wider applicability, a nonperturbative analysis is desirable.
A powerful tool to study nonperturbative aspects of \ac{QFT} is the use of lattice field theory, where the \ac{QFT} is seen as the description of the long-distance limit of a quantum theory defined on a discrete spacetime (or spatial) lattice at its quantum critical point.
Currently, the only way to do computations with 4d lattice field theory is with \ac{MC} methods, but
the inherent nonequilibrium nature of the entanglement harvesting protocols means that even the lattice \ac{MC} methods fail with severe sign problems. An alternative approach to conventional lattice field theory and its simulation is needed. 

In this work, we take a first step towards a nonperturbative study of entanglement harvesting in \ac{QFT} using lattice field theory. 
The question we are interested in this work is: can one distill Bell pairs out of the ground state of a relativistic \ac{QFT}, and, if so, at what rate?
In other words, how can two parties---conventionally called Alice and Bob---do cooperative quantum tasks using the vacuum?
As discussed above, this is a difficult question to probe quantitatively directly in the continuum for relativistic \acp{QFT}.
However, in low dimensions, the quantum systems whose second-order critical points are described by the continuum \acp{QFT} allow the use of tensor network techniques in a study of entanglement and their real-time dynamics.
More specifically, we focus on one of the simplest lattice field theories, the one-dimensional \ac{TFIM}, and analyze a simple protocol for the distillation of Bell pairs out of the ground state (or vacuum) of the theory using tensor networks.

The number of entangled Bell pairs which can be extracted from a system can be thought of as a measure of entanglement, called the ``entanglement of distillation'' \cite{GUHNE20091}, which is however difficult to compute. It has been shown that entanglement negativity
\cite{vidal_computable_2002} gives an upper bound to distillable entanglement, and thus can be a useful proxy for it.
Gaussian states have been used to compute negativities in free (bosonic and fermionic) lattice field theories
\cite{shapourian_partial_2017a, shapourian_finitetemperature_2019, cornfeld_measuring_2019, chang_entanglement_2016, audenaert_entanglement_2002a} and study their continuum limits \cite{PhysRevD.103.065007,Klco:2021biu,Klco:2021cxq}. However, these methods do not work for interacting field theories.
In 1+1 dimensions, \ac{CFT} methods have been used to compute negativities
\cite{ruggiero_negativity_2016, chang_entanglement_2016, calabrese_finite_2014, calabrese_entanglement_2013a, calabrese_entanglement_2012, audenaert_entanglement_2002a}, and even the realtime dynamics after local quenches in some cases
\cite{wen_entanglement_2015a, hoogeveen_entanglement_2015, eisler_entanglement_2014}.
However, the computation of the negativity between two finite and disjoint intervals, which is the one relevant to us, is still difficult in general.  

On the other hand, given a quantum state represented as a \ac{MPS}, the entanglement negativity between two disjoint regions \emph{can} be computed efficiently.
Therefore, for low-dimensional (1d or 2d) systems, tensor networks provide a powerful method for computing entanglement negativity, which we exploit in this work.

This paper is organized as follows. In \cref{sec:negativity}, we provide a brief pedagogical summary of the basic concepts related to entanglement negativity and entanglement harvesting.  In \cref{sec:groundstate}, we define entanglement negativity of the ground state of the \ac{TFIM} and investigate its behavior across the phase transition. In \cref{sec:singleswap}, we introduce a simple protocol for entanglement harvesting from a quantum spin chain and study the protocol 
in the thermodynamic limit,
and describe our results for the \ac{TFIM}. We present our conclusions in \cref{sec:conclusions}.
We explain a few technical details in the appendices. \Cref{app:effdens} is a pedagogical introduction to the use of reduced density matrices to single quantum systems; \cref{app:perm} analyzes the role of permutation symmetry in studying density matrices describing systems drawn from an ensemble; and \cref{sec:rdmbp} discusses the use of reduced density matrices for a thermodynamic system where ergodicity is broken. \Cref{app:mps} describes the \ac{MPS} ansatz and its implementation in our work, and in \cref{app:exact},  results from some exact calculations are compared against those obtained using the \ac{DMRG}. Finally, \cref{app:align} describes our parameterization of maximally entangled 2-qubit states.


%% file: sec_negativity.tex
\section{Entanglement Negativity}
\label{sec:negativity}

There are several quantities in quantum mechanics that allow us to assess the quantum nature of the physical system. The most well known example is entanglement. If we consider some quantum system $S$ in a pure state, but focus only on some subsystem $A$ so that we only have access to information within $A$, then we can learn how $A$ is entangled with $S$ through the reduced density matrix\footnote{In \cref{app:effdens}, we briefly review some basics of density matrices as we use in our work.} $\rho^{\rm r}_A$. For example, the von Neumann entropy
\begin{align}
S_A \ =\ -\mathrm{Tr}_A\left(\rho^{\rm r}_A \ln\rho^{\rm r}_A\right),     
\end{align}
is nonzero if $A$ is entangled with the remaining system and, thus, contrary to classical expectations, we will find $A$ to be in a mixed state, even though the density matrix $\rho$ of the full system $S$ describing a pure state has vanishing von Neumann entropy.

If instead of a single subsystem we consider two subsystems $A$ and $B$ inside some bigger quantum system $S$, the situation is more complex. As before, let us assume\footnote{It is well known that given a system in any state, one can always find a pure state of a hypothetical bigger system \(S\) from which the given state is obtained as a reduced density matrix.} the full system $S$ is in some pure state so that the density matrix $\rho$ of the full system has a vanishing von Neumann entropy. For a generic quantum many body state, the combined subsystem $A$ and $B$ will be in a mixed state and entangled with the rest of the system. This means that reduced density matrix $\rho^{\rm r}_{AB}$ will, in general, be characterized by a nonzero von Neumann entropy. For such mixed states, the easiest way to detect entanglement\footnote{Standard classical measures like mutual information do not discriminate between classical and quantum correlations, both of which can be nonzero if the combined system $AB$ is not pure.} between $A$ and $B$ is by using an ``entanglement witness,'' an operator that is positive semidefinite when restricted to unentangled states, but not in general. (See Ref.~\cite{GUHNE20091} and references therein.)
One such class of operators arises from the positive maps on one subsystem (say $A$) but that are not positive on the entire system when combined with the identity operation on the other subsystem. These are called positive but not completely positive maps.

A convenient entanglement witness in this class is the hermiticity and trace-preserving `partial transpose'~\cite{PhysRevLett.77.1413}. It is easiest to define this operator using a direct-product basis for the joint state of the subsystems $A$ and $B$, i.e., a basis using states $\ket{ab} \equiv {\ket a}_A \otimes {\ket b}_B$, built out of any orthonormal basis, $\{{\ket a}_A\}$ and $\{{\ket b}_B\}$, for the Hilbert spaces of subsystems $A$ and $B$ respectively. Then we can express the reduced density in terms of its components $\rho^{ab}_{a'b'}$ as 
\begin{align}
  \rho_{AB} &\equiv \sum_{\substack{a,b\\a',b'}} \rho^{ab}_{a'b'} \ket{ab} \bra{a'b'},
\end{align}
which is clearly an operator in the direct product space of $A$ and $B$. The hermiticity, positivity, and trace-preserving `partial transpose' operator can then be defined as $(T_A\otimes I_B)$ whose action on $\rho_{AB}$ is defined as
\begin{align}
\rho_{AB}^{T_{A}} \equiv (T_A\otimes \Id_B)\ \rho_{AB} 
                  \equiv \sum_{\substack{a,b\\a',b'}} \rho^{a'b}_{ab'} \ket{ab} \bra{a'b'}\,.
\end{align}
Note that though the partial transpose operation depends on the basis chosen, the operations defined with respect to different orthonormal bases for each of the subsystems are all unitarily equivalent, and so $\rho_{AB}^{T_{A}}$ has the same eigenvalue spectrum in all bases. In particular the trace norm\footnote{Conventionally, \(\sqrt{x^\dagger x}\) stands for the absolute value of \(x\). We also extend all functions of real variables to self-adjoint matrices via eigendecomposition.} 
\begin{align}
\mathopen{||}\rho^{T_A}_{AB}\mathclose{||}_1\equiv\Tr\sqrt{(\rho^{T_A}_{AB})^\dagger\rho^{T_A}_{AB}}   
\end{align}
is basis independent. Further, it is easy to verify that $\mathopen{||}\rho^{T_B}_{AB}\mathclose{||}_1 = \mathopen{||}\rho^{T_A}_{AB}\mathclose{||}_1$.

Entanglement negativity ${\cal N}(\rho_{AB})$ between two subsystems $A$ and $B$ characterized by the density matrix $\rho_{AB}$ is then defined as the sum over all negative eigenvalues of $\rho_{AB}^{T_{A}}$~\cite{vidal_computable_2002}, which can be conveniently computed through the relation
\begin{align}
  {\cal N}(\rho_{AB}) \equiv \frac{\mathopen{||}\rho^{T_A}_{AB}\mathclose{||}_1 - 1}{2}.
  \label{eq:neg-defn}
\end{align}
The logarithmic negativity is then defined through the relation
\begin{align}
  E_N(\rho) \equiv \log_2 (\mathopen{||}\rho^{T_A}\mathclose{||}_1) = \log_2(2{\cal N}+1).
  \label{eq:logneg-defn}
\end{align}
Any density matrix $\rho_{AB}$ with a positive-semidefinite partial transpose has ${\cal N}(\rho_{AB}) = 0$ and the system $AB$ is said to be in a \ac{PPT} state. Otherwise, ${\cal N}(\rho) > 0$, and we say that $AB$ is in a \ac{NPT} state.

The importance of entanglement negativity arises from the fact that separability (i.e., lack of quantum entanglement) implies \ac{PPT}, therefore
all \ac{NPT} states are entangled.
The converse is not generally true.
It can be shown that ${\cal N}(\rho_{AB}) = 0$ is a necessary and sufficient condition for the state to be separable
only when the Hilbert space dimension of the two subsystems are $2\times 2$ or $2\times 3$ ~\cite{horodecki_violating_1995}. In other words, for general Hilbert space dimensions there are states with ${\cal N}(\rho_{AB}) = 0$, which are nonetheless entangled\footnotepunct{{For \(2\times2\) and \(2\times3\), any positive map on \(A\) can be decomposed as \(\phi_A + T_A \psi_A\), where \(\phi_A\) and \(\psi_A\) are completely positive maps on density matrices and $T_A$ is the partial transpose operation. For higher dimensions, the set of independent nondecomposable positive maps is uncountable; as a result, there are other independent entanglement witnesses, and the negativity criterion for detecting entanglement does not generalize~\cite{doi:10.1142/S1230161211000224,doi:10.1063/1.4966984}.}}.

Quantum entanglement is a resource than can be used to carry out a variety of tasks not possible with access to classical correlations alone~\cite{nielsen00}. However, not all entangled states are equally useful. In fact, most quantum algorithms rely on the availability of maximally entangled qubits, which are called Bell states. Entanglement harvesting is a method or a protocol for extracting Bell states from the quantum system $S$. For example, suppose Alice and Bob are two observers who have access only to their own subsytems $A$ and $B$ within the full system $S$. They can perform local measurements on their own subsystems by acting with local completely positive maps\footnote{These include all physically realizable local operations: including unitary operations on their subsystem and ancillae, as well as projective measurements and postselection.} on their subsystem. As explained in \cref{app:effdens}, the reduced density matrix $\rho_{AB}$ contains all the information accessible to Alice and Bob, and any question they can ask about the system can be formulated in terms of $\rho_{AB}$. Let us further assume Alice and Bob can communicate with each other classically about the results of their measurements.
The set of all local operations that Alice and Bob can make, possibly dependent on results communicated by the other, is collectively referred to as \ac{LOCC}. An important property of \ac{LOCC} operations is that though they can increase the classical correlations between the subsystems, they cannot increase quantum entanglement between them.

A natural question one may ask is, given many independent copies of subsystem $AB$, each in the same mixed entangled state $\rho_{AB}$, can one observer\footnote{Here we assume that both Alice and Bob know the state \(\rho_{AB}\). If it is known that the copies are independent and identical, and one can use an unlimited number of them, then LOCC operations can be used to determine this state to arbitrary accuracy, so this is not an important assumption. See~\cref{app:perm} for a discussion of the more general case.} who has access only to $A$ (Alice) and another observer who has access only to $B$ (Bob) produce pure Bell pairs using only \ac{LOCC} operations. If this can be done, then the state $\rho_{AB}$ is called \emph{distillable}. Schematically we can write this distillation process through the equation
\begin{align}
  \underbrace{\bigotimes_n \rho_{AB}}_{n \text{ copies}} \xrightarrow{\text{distillation}}  \underbrace{\bigotimes_m \frac{1}{\sqrt{2}} \left( \ket{\uparrow\downarrow} - \ket{\downarrow\uparrow} \right)}_{m \text{ copies}} \otimes \underbrace{\rho_{\rm discarded}}_{n-m \text{\ discards}},
  \label{eq:distillation}
\end{align}
where \(\bigotimes_n \rho\) means a direct product of \(n\) copies of \(\rho\). We will discuss a particular distillation protocol later in \cref{sec:prot-multi}. It has been shown that \ac{PPT}
states are not distillable~\cite{horodecki_mixedstate_1998}. Using this idea, we can define ``entanglement of  distillation" as the expected asymptotic number ($\lim_{n\to\infty} m/n$) of Bell pairs that can be distilled per copy of the initial system~\cite{vidal_computable_2002}. It can be shown that the logarithmic negativity $E_N(\rho)$, defined in \cref{eq:logneg-defn}, is an upper bound on the entanglement of distillation
\footnotepunct{Since \ac{PPT} does not imply separability, this means that there are \ac{PPT} states, which are entangled, but even from many copies of which one cannot extract any Bell-pairs. Such states are called \emph{bound entangled}. Though they cannot be distilled, they can, however, be used to increase the entanglement in distillable states~\cite{Horodecki:1998vm}.}.

The question we are interested in this work is: can one distill Bell pairs out of the ground state of a relativistic \ac{QFT}, and, if so, at what rate? In other words, how can Alice and Bob do cooperative quantum tasks using the vacuum?
This is a difficult question to probe quantiatively, since ground states of relativistic \acp{QFT} are not easily accessible beyond perturbation theory.
Further, any protocol of distillation of Bell pairs requires studying realtime dynamics, which cannot be studied for lattice field theories with typical \ac{MC} methods.
However, continuum \ac{QFT} arise near a second-order critical point of quantum spin chains, and this allows us to use tensor network techniques to study entanglement and realtime dynamics of low-dimensional continuum \acp{QFT}.
In this work we focus on the simplest lattice field theory, the \ac{TFIM}, and analyze a simple protocol for the distillation of Bell pairs out of the ground state (or vacuum) of the theory. The study of entanglement negativity in the ground state is a natural starting point of our discussion, which we consider in the next section. 


%% file: sec_groundstate.tex
\section{Entanglement Negativity in the ground state of the Ising model}
\label{sec:groundstate}

\begin{figure}[htbp]
  \centering
  \includegraphics[width=0.7\linewidth]{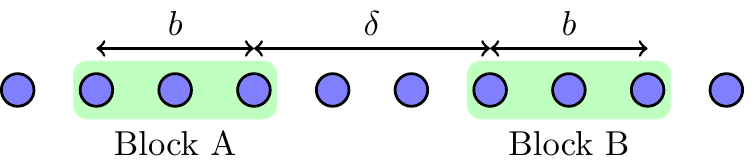}
  \caption{We compute the entanglement negativity ${\cal N}(b, \delta)$ for two blocks $A$ and $B$ consisting of $b$ sites each and separated by $\delta$ sites. }
  \label{fig:neg-blocks}
\end{figure}

\begin{figure*}[tb]
\centering
\includegraphics[width=0.75\linewidth]{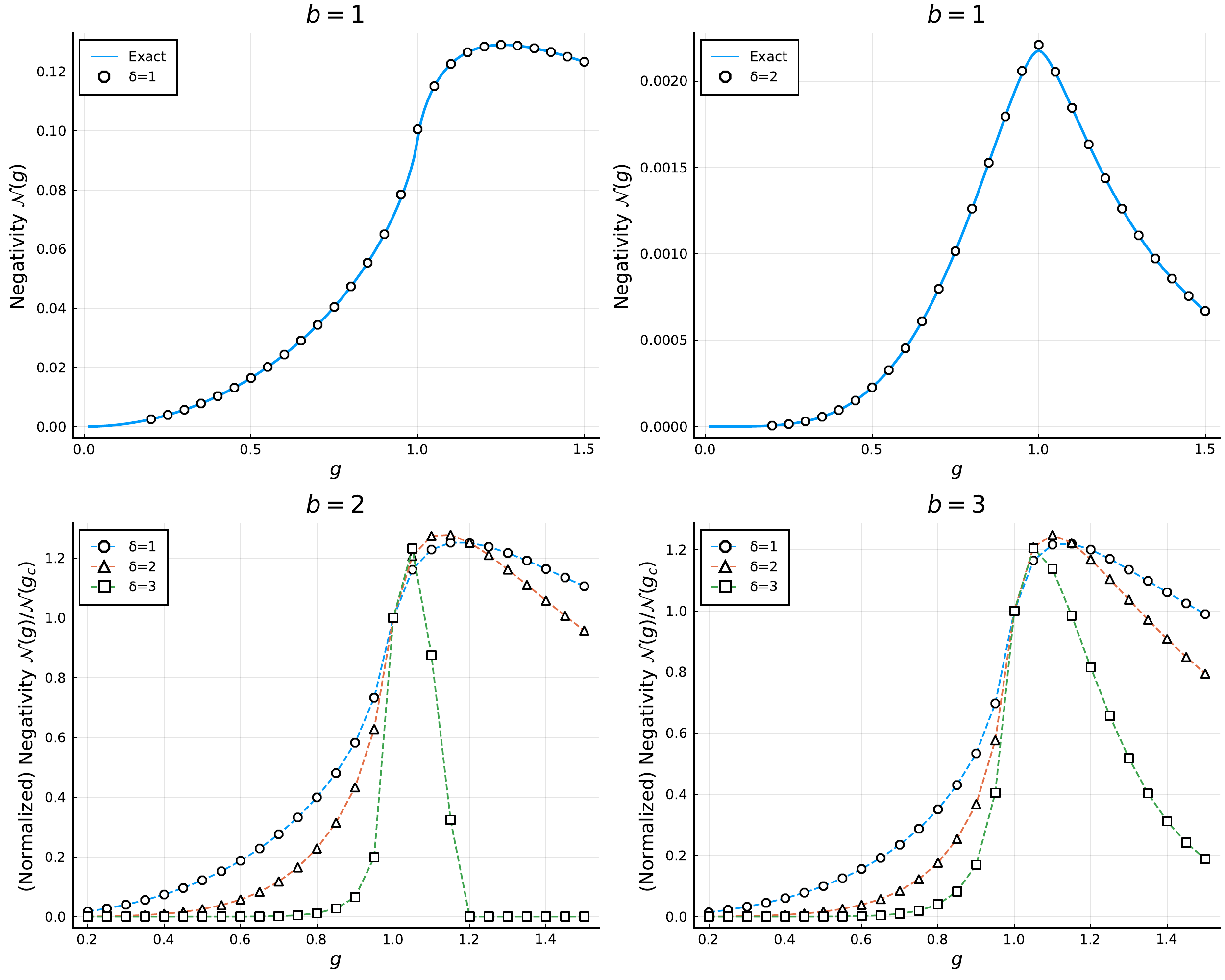}
\caption{Entanglement negativity for lattice length $L=64$, blocksizes $b=1,2,3$ and seperations $\delta=1,2,3$ as function of the coupling $g$. The \emph{Top row} shows results for $b=1$. The top-left plot shows nearest-neighbor negativity ($\delta=1$) and the top-right plot shows next-to-nearest neighbor negativity ($\delta=2$). (Negativity for $\delta=3$ is zero for $b=1$.) Since the negativities with $b=1$ can be  computed exactly for the \ac{TFIM} (see \cref{app:exact}), we also show the exact results as thick solid lines. \emph{Bottom row:} The left panel shows blocksize of $b=2$ while the right panel shows $b=3$ for various block separations $\delta = 1,2,3$. Note that here we normalize the curves by $\mathcal{N}(g=1.0)$ to fit the various separations in the same figure. In all these cases (except $b=1, \delta=2$), we find that the maximum negativity is achieved slightly to the right of the critical point.}
\label{fig:neg-transition}
\end{figure*}

The lattice system of interest in our work is a one dimensional open chain of $L$ quantum spin-half particles interacting with the Hamiltonian which is the well known \ac{TFIM}~\cite{*[{See review }] [{ and references therein.}] Stinchcombe_1973}. We label the sites of the spin chain as $i=1,2,...,L$ so that the Hamiltonian is given by 
\begin{align}
 H = -\sum_{i=1}^{L-1} Z_{i}Z_{i+1} - g \sum_{i=1}^{L} X_i,
  \label{eq:H-tfim}
\end{align}
where $X_i,Y_i,Z_i$ are the usual Pauli sigma matrices associated to the site $i$ and $g$ is the coupling to the transverse field. Note that the spin chain has \ac{OBC}. In the thermodynamic limit $(L\rightarrow \infty$) this model is known to have a second-order quantum phase transition at $g = 1$ between an ordered phase ($g<1$) and a disordered phase ($g>1$). The local order parameter of the theory is the expectation value $\langle Z_i\rangle$ of the $z$-component of the quantum spin, and is related to the $\ZZ_2$ Ising symmetry of $H$ with the generator
\begin{align}
Q = \prod_{i=1}^{L} X_i.
\end{align}
The ground state of the Hamiltonian \cref{eq:H-tfim} can be written as
\begin{align}
\ket{\Omega(g)}\ =\ \sum_{\sigma_1, \dotsc, \sigma_{L}} \Psi(\sigma_1,\sigma_2,...,\sigma_L)\ | \sigma_1 \sigma_2 \cdots \sigma_{L} \>
\label{eq:gst-0}
\end{align}
where $\sigma_i$ label some suitable eigenvalues of the local quantum spins. For example at extremely large $g$, the transverse-field term dominates and the (unique) ground state is just a product state of local eigenstates of $X_i$:
\begin{align}
  \ket{\Omega(g\to\infty)} = \ket{\rightarrow \cdots \rightarrow},
\end{align}
where $\ket{\rightarrow}$ is the local eigenstate of $X$ with eigenvalue $1$. At the other extreme, when $g = 0$, the ground state is doubly degenerate and the two basis states of this degenerate subspace can be chosen to be
\begin{align}
  \ket{\Omega(0)^+} &=  \ket{\uparrow \cdots \uparrow},\\
  \ket{\Omega(0)^-} &= \ket{\downarrow\cdots \downarrow}.
                 \label{eq:ground-states}
\end{align}
Note that $Q|{\Omega(0)^+}\rangle = \ket{\Omega(0)^-}$.
At some intermediate coupling the ground state is more complicated but can still be solved exactly for any value of the coupling $g$ \cite{Sachdev:QPT}. Such a calculation uses the nonlocal Jordan-Wigner transformation to map the model into a free fermion model which can then be used to compute the ground state.
For $g<1$, the ground state is doubly degenerate due to spontaneous symmetry breaking.
Since we are interested in harvesting the entanglement of the ground state, we need to pick one of the two states in \cref{eq:ground-states}.
To ensure that we always work with $|\Omega^{+}\>$ when $g<0$, we add a small pinning field at the boundary in our numerical calculations\footnotepunct{See \cref{sec:rdmbp} for a technical discussion of this issue and the smallness criteria and \cref{app:mps} for details on the numerical methods.}.
Adding this pinning field then defines the ground state uniquely for all values of the coupling $g$.

 \begin{figure*}[tbp]
\centering
\includegraphics[width=0.31\linewidth]{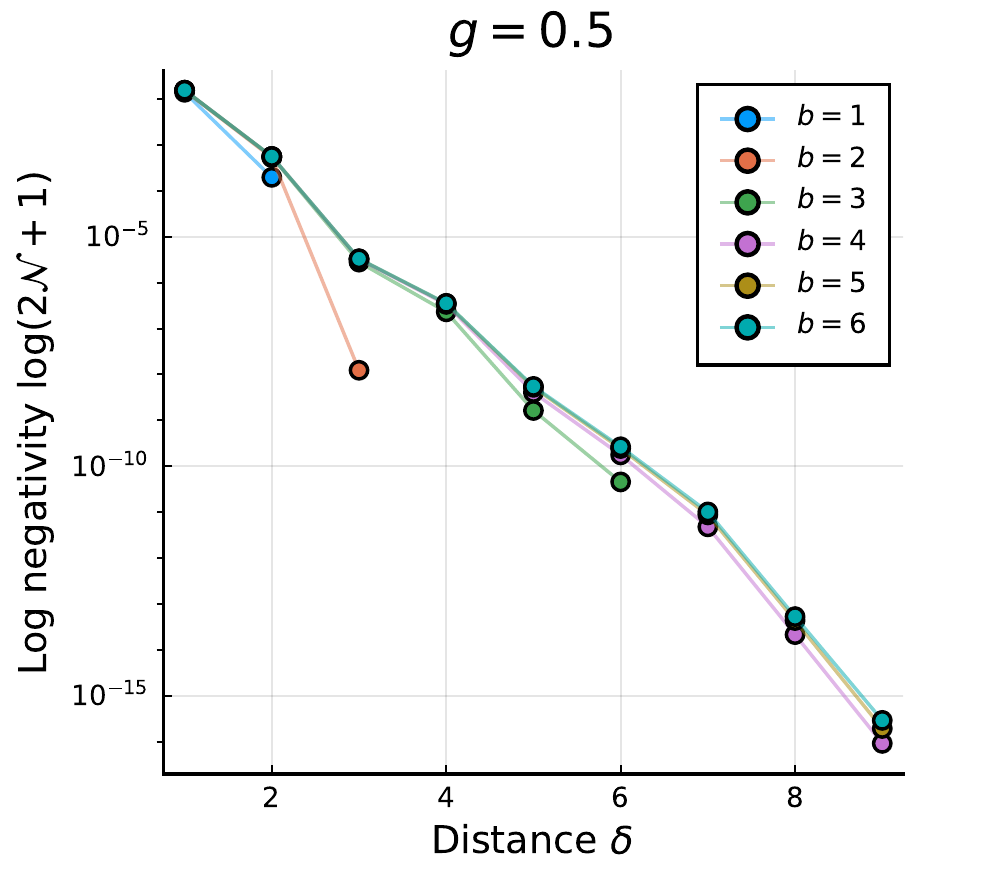}
\includegraphics[width=0.31\linewidth]{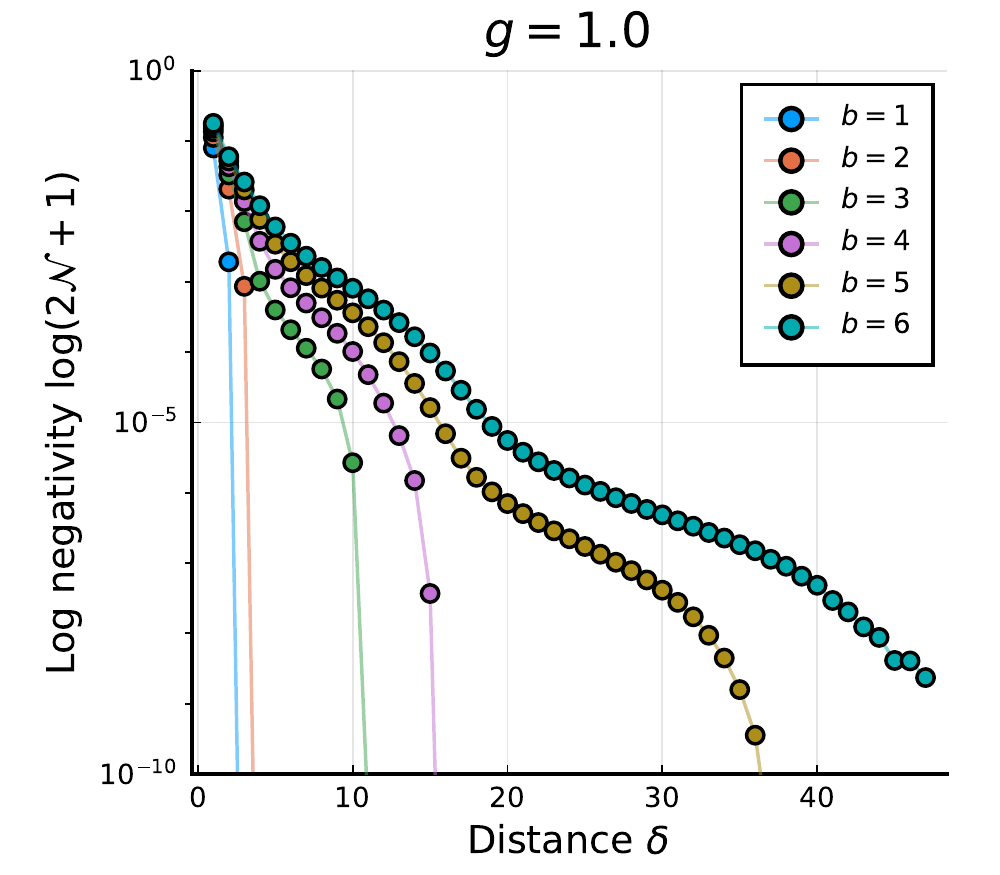}
\includegraphics[width=0.31\linewidth]{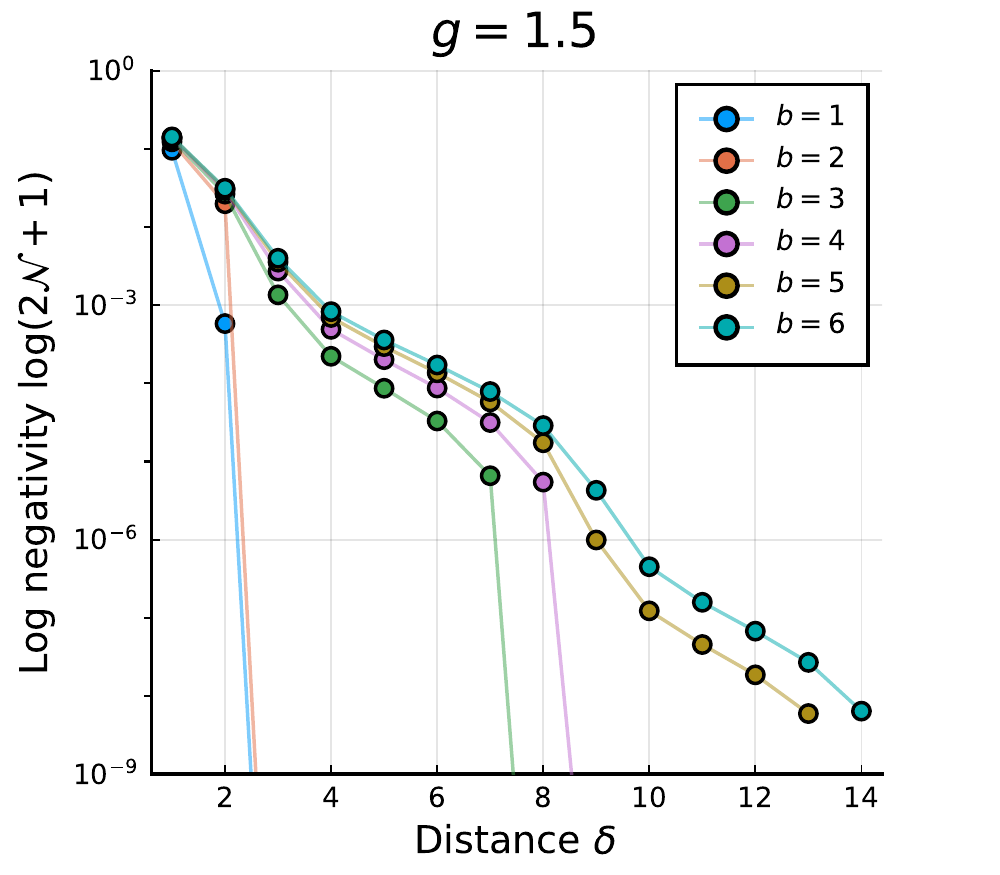}
\caption{\label{fig:death}
  Sudden death of entanglement in the transverse field Ising model for various values of $g$. The various lines show the entanglement negativity as a function of the separation $\delta$ for different block sizes $b$. For small $b$, the negativity drops quickly to zero beyond some separation $\deltaMax(b)$. As we increase the block size, $\deltaMax$ becomes larger and we can separate the regions further while still having nonzero negativity. As we approach larger $b$, the decay is exponential with $\delta$, as expected from the literature. For $g<1$, the decay seems to be described roughly by an exponential function with a small wiggle. For $g\geq 1$ we notice that the wiggles turn into smoother oscillations.
}
\end{figure*}

Our goal is to compute the entanglement negativity of the ground state between two spatial regions $A$ and $B$ in the \ac{TFIM} as we discussed in the previous section. To define the negativity, we first note that the Hilbert space $\HSp$ of the full system $S$ factorizes into a tensor product of the local Hilbert spaces $\HSp_i$ at each lattice site $i\in S$, such that $  \HSp = \otimes_{i\in S} \HSp_{i}$.
We can identify two nonoverlapping set of sites, one of which belongs to $A$ (Alice) and the other to $B$ (Bob) and call the rest of the sites as $C$. The spins in each of the regions $A$ and $B$ will be assumed to be spatially connected but the two regions will be separated from each other. In \cref{fig:neg-blocks} we show an illustration of such regions, where each region ($A$ or $B$) consists of $b$ sites and the two regions are separated by $\delta$ sites. We can express the ground state \cref{eq:gst-0} as
\begin{align}
\ket{\Omega(g)}\ =\ \sum_{\sgA,\sgB,\sgC} \Psi(\sgA,\sgB,\sgC) |\sgA,\sgB,\sgC\>
\end{align}
where $\sigma^{A}, \sigma^{B}, \sigma^{C}$ denote the collective spins in regions A,B,C, respectively.
For example, $\sgA$ labels one of $2^b$ spin configurations within the region A and so on. 
The \emph{reduced} density matrix of the subsystems $A\cup B$ can then be defined as
\begin{align}
  \rho^{\rm r}_{AB} = \Tr_{C} \rho\,.
  \label{eq:reddensmat}
\end{align}
where trace is over all spin degrees of freedom $\sgC$ in the region C that do not belong to either $A$ or $B$, and
\begin{align}
  \rho  = | \Omega(g)  \> \< \Omega(g) |\,.
\end{align}
It is then possible to compute the entanglement negativity \({\cal N}(b,\delta;g)\) using \cref{eq:neg-defn} and understand how it depends on the block size, \(b\) (assumed same for both), the separation between the blocks, \(\delta\), and the coupling \(g\) (See~\cref{fig:neg-blocks}).
Since our calculations are performed on a finite lattice with \ac{OBC}, we expect \({\cal N}(b,\delta;g)\) will also depend on the lattice size $L$ and where the two regions $A$ and $B$ are located with respect to the boundaries. To minimize such boundary contributions, we place $A$ and $B$ symmetrically on either side of the center of the quantum spin chain. Although the \ac{TFIM} can be solved exactly, it is difficult to compute ${\cal N}(b,\delta;g)$ analytically.
For this reason, we use \ac{MPS} for our calculations.
Some of the details of how our calculations are performed is explained in \cref{app:mps}. 

In \cref{fig:neg-transition}, we show the behaviour of the negativity ${\cal N}(b,\delta;g)$ as defined in \cref{eq:logneg-defn} as a function of the coupling $g$ for a few values of seperation \(\delta\) and block sizes \(b\). For the block size $b=1$, we also compare against the exact results, explained briefly in \cref{app:exact}. At both extremes, $g=0$ and $g\to \infty$, the ground state is described by a product state and therefore the negativity vanishes as seen in our plots. It is well known that several entanglement measures display a qualitative change across a phase transition \cite{osborne_entanglement_2002, osterloh_scaling_2002}. The next-to-nearest-neighbor $b=1, \delta=2$ negativity ${\cal N}(1, 2; g)$ shows a clear peak at the critical point $g=1$.
However, the other negativities are not maximized exactly at the critical point, but for slightly larger $g$.  From the point of view of this work, therefore, we would like the spin-chain to be close to criticality to extract the maximum amount of entanglement.  For a condensed-matter system we can experimentally tune $g$ close to a transition, while for discussing continuum \ac{QFT}, we need to take the continuum limit which also arises close to the critical point.
Unfortunately, the tensor network computations become difficult when correlation length diverges. Therefore, we choose a range of values around $g=g_c$ to perform our computations of the harvesting protocol.

The long distance physics of quantum many-body systems is described by quantum field theories\footnotepunct{The Hilbert space of a continuum field theory is not a tensor product of `local Hilbert spaces\rlap', but do obey the split property~\cite{Roos1970,Buchholz1974}. On the other hand the Hilbert space of a lattice field theory is a tensor product of local Hilbert spaces. Near the continuum limit, the correlation length diverges, and a continuum field theory describes the low-energy physics, appropriately scaled by the correlation length. This low-energy truncated Hilbert space is not describable as a direct product but the split property holds at nonzero scaled distances. Since we only consider the negativity between separated regions on the lattice, without an explicit truncation of the Hilbert space, we will not worry about these subtleties of the continuum in this work.}. For critical systems, an analysis of negativity in conformal quantum field theories~\cite{alba_entanglement_2013, calabrese_entanglement_2012, calabrese_entanglement_2013a} shows that the logarithmic negativity $E_N$ decays exponentially: $E_N(b, \delta; g) \sim \exp(-c\,\delta/b)$ for some exponent $c$ and large enough $\delta$ and $b$ with $\delta/b$ finite. However, this negativity suddenly drops to zero as we increase the separation $\delta$, keeping $b$ and $g$ fixed, beyond some finite $\delta = \delta_{\text{max}}$~\cite{*[{See review }][{ and references therein}]doi:10.1126/science.1167343}.
We show our results for the exponential decay and this drop (called the sudden death of entanglement) in \cref{fig:death}.
For extracting negativities out of a finite spin system, this means that the separation is bounded by $\delta_{\text{max}}$. To take the continuum limit of our protocol, we would therefore have to increase both block sizes $b$ and separation $\delta$ while keeping $\delta/b$ fixed. Larger $b$ makes both the time-evolution and negativity computations harder. For this prototypical study, we set $b=1$ and hence are limited to $\delta=1$.


%% file: sec_protocol_single.tex
\begin{figure*}[htbp]
  \centering
 \includegraphics[width=0.99\linewidth]{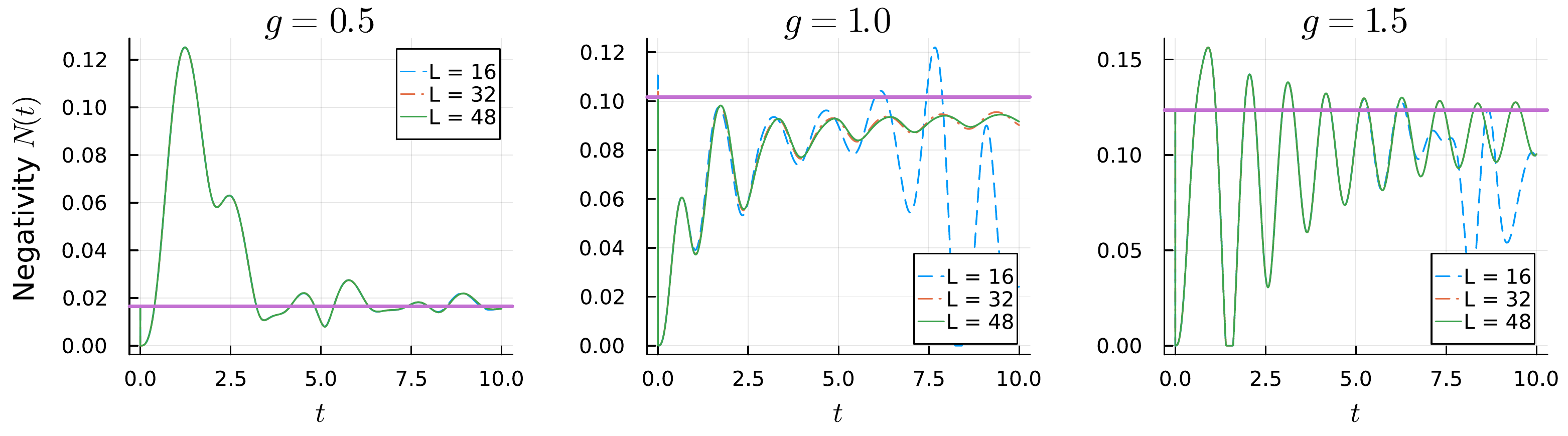}
 \caption{\label{fig:time-evolution}
Here we show the time evolution of two-point negativity ${\cal N}_1(t)$ as a function of $t$ for various lattice sizes at different values of the coupling $g$, corresponding to $g=0.5$ (left,ordered), $g=1$ (center,critical) and $g=2$ (right,disordered) phases.
The units of negativity are such that a bell-pair has negativity $\Neg=1$. The time is in units of inverse energy. We perform time evolution in steps of $\epsilon=0.01$ and perform $5000$ time steps. The solid horizontal line is the negativity ${\cal N}_0$
of the ground state of the \ac{TFIM} which is also the negativity obtained from $\rho^R_0$ after the first swap into the register $\alpha=0$. After this initial swap, we start measuring ${\cal N}_1(t)$ obtained from $\tilde{\rho}^R_1$ and so this negativity immediately drops to zero because at $t=0$ the spins were just swapped into the system. Then the negativity begins to grow as the state evolves according to the Hamiltonian of the the \ac{TFIM}. We note that there is a well-defined thermodynamic limit which is reached with $L=32$ in the range of times shown. Finite size effects cause fluctuations in the thermodynamic curve at large times, which are clearly visible in the $g=1.0$ and $g=1.5$ plots. In the thermodynamic limit, after a long time, the negativity ${\cal N}_1(t)$ goes back to its equilibrium value of ${\cal N}_0$ as expected. Fortunately, we can optimize the extracted negativity by using the first peak instead of the equilibrium value ${\cal N}_0$ which occurs at short times. For this we can work with smaller system sizes and in our work we use $L=16$ in all of our calculations.}
\end{figure*}

\section{Entanglement Harvesting Protocol}
\label{sec:singleswap}

We now explain the protocol that we use to harvest entanglement from the spin chain of the \ac{TFIM} into two registers one of which is with Alice and the other with Bob. Each of these registers contains $N$ sets of $b$ quantum spins, the same size as $\sgA$ and $\sgB$, that are local to Alice or Bob. Let us label each set of $b$-spins as ${\sA}_i$ and $\sB_i$ where $i=0,1,2...,N-1$. The goal of our protocol is to create an entangled state of these two registers containing $2bN$ spins, by allowing them to temporarily interact with the spins $\sgA$ and $\sgB$ in the spin-chain of the \ac{TFIM}. These interactions can be viewed as discrete operations performed by Alice and Bob on the spin-chain and their registers. One particularly easy operation that Alice can perform is the unitary process $U^A_\alpha$ that `swaps' the $b$ spins ${\sA}_\alpha$ that she has, with the $b$ spins $\sgA$ in the \ac{TFIM}. This process can be defined through the relation
\begin{align}
U^A_\alpha & |\sA_1,..\sA_\alpha,..\sA_N \rangle \otimes |\sB_1,..\sB_N\rangle \otimes |\sgA,\sgB,\sgC\rangle \nonumber \\
& = |\sA_1,..\sgA,..\sA_N\alpha\rangle \otimes |\sB_1,..\sB_N\rangle \otimes |\sA,\sgB,\sgC\rangle
\label{eq:swap}
\end{align}
Note that the Hilbert space basis now involves not only the spins of the \ac{TFIM}, but also those in the two registers, each with $bN$ spins. At the end of the swap process the wave function of the combined system  $\psi(\sA_1,..\sA_\alpha,..;\sB_1,..;\sgA,\sgB,\sgC)$ changes to $\psi(\sA_1,..\sgA,..;\sB_1,..;\sA_\alpha,\sgB,\sgC)$ which means that if the wavefunction is written as a matrix in the product basis, the swap simply transposes the matrix in the relevant indices. Bob can perform a similar measurement through the process \(U^B_\alpha\) which swaps $\sB_\alpha$ and $\sgB$. The two swaps commute: $U^A_\alpha U^B_\alpha = U^B_\alpha U^A_\alpha$.

In order to distill entanglement from the spin-chain into the registers, in our protocol Alice and Bob both begin with un-entangled spins in their registers and perform the swap $U^A_0U^B_0$ between their region of the spin-chain and the $\alpha=0$ set of $b$-spins of their registers. After the swap they allow the spin-chain to evolve according the Hamiltonian of \ac{TFIM} (i.e., without further interaction with the registers) for some time and again perform the swap, except this time they swap their respective regions of the spin chain with the $\alpha=1$ set of $b$-spins of their registers. This process continues and in general, at the \(\alpha\)th step, they perform the swap $U^A_{\alpha}U^B_{\alpha}$ between their region of the spin-chain and the $\alpha$th set of $b$-spins of their registers. The entanglement properties that are distilled from the spin-chain can be studied using the reduced density matrix of the registers $\rho^R_\alpha$ after swapping the $\alpha$th set of $b$-spins in the registers. Note that $\rho^R_\alpha$ is a $2^{2b(\alpha+1)}\times 2^{2b(\alpha+1)}$ matrix. As we explain below, it is useful to discard the $\alpha=0$ spins of the registers (trace over them) and define $\tilde{\rho}^R_\alpha$ which is a $2^{2b\alpha}\times 2^{2b\alpha}$ matrix.

\subsection{Results: Single step}

As mentioned earler, the amount of entanglement that can be distilled out of the registers is bounded by the logarithmic negativity of $\tilde{\rho}^R_\alpha$, but we can build up this negativity by making $\alpha$ large (i.e., repeat the swap process several times). However, the timings when the swaps are performed can also play a role. In order to understand how to efficiently build up the negativity in the registers, we now study how it builds up in the spin-chain as it evolves following a swap. To this end, we study the simple case of $b=1$ and $\delta=1$. We begin with the spin-chain in the ground state of the \ac{TFIM} and perform the first swap ($U^A_0U^B_0$) at time $t=0$. Let us define ${\cal N}_0$, as the negativity obtained from $\rho^R_0$ after this swap. This is the negativity between the swapped spins in the ground state of the \ac{TFIM}. After this initial swap let us allow the spin-chain to evolve according to \cref{eq:H-tfim} for a time $t$ before performing the second swap ($U^A_1U^B_1$). After this second swap we can compute the density matrix $\tilde{\rho}^R_1(t)$, and the associated negativity ${\cal N}_1(t)$ as discussed in \cref{sec:negativity}. In \cref{fig:time-evolution}, we plot the time evolution of ${\cal N}_1(t)$ in the three phases as a function of the system size $L$. Note that ${\cal N}_1(t)$ vanishes at $t=0$ since the spins swapped into the spin chain have not yet had the time to evolve, but then begins to build up. Further, it is described by a well defined function of $t$ in the thermodynamic limit, which starts at the origin as expected and approaches ${\cal N}_0$ in the limit $t \rightarrow \infty$. At intermediate times it oscillates and reaches a maximum at some $t=t_{\rm min}(g)$, where ${\cal N}_1(t_{\rm min})> {\cal N}_0$ in both the ordered and disordered phases. 


%% file: sec_protocol_repeat.tex
\begin{table*}
\input{tables/tab_fidelities}
\caption{Fidelity $F(\bar{\rho}_i,\bar{\rho}_j)$ between different $\bar{\rho}_i$'s  where $i=0,1,2,3,4$. Each $\bar{\rho}_i$ is a $4\times 4$ matrix of the $i$th spin-pair obtained by tracing $\rho^R_{\alpha=4}$ over the remaining spin-pairs. }
\label{tab:fid1} 
\end{table*}

\begin{table*}
\input{tables/tab2.tex}
\caption{\label{tab:fid2} Comparison of various density matrices discussed in the text using Fidelity $f_{X,Y} = f(\rho^X_\alpha,\rho^Y_\alpha)$ and $\tilde{f}_{X,Y} = f(\tilde{\rho}^X_\alpha,\tilde{\rho}^Y_\alpha)$, defined using the relation \cref{eq:fidk}.}
\end{table*}

\begin{table*}[htbp]
\renewcommand{\arraystretch}{1.0}
\input{tables/tab3}\\
\caption{The log-negativity defined through $E_1 = E_N(\bar{\rho}^R_\alpha)$,  $E_2 = E_N(\rho^R_\alpha)/(\alpha+1)$, $E_3 = E_N(\bar{\rho}^S_\alpha)$, $\tilde{E}_2 = E_N(\tilde{\rho}^R_\alpha)/\alpha$, $\tilde{E}_3 = E_N(\bar{\tilde{\rho}}^S_\alpha)$ where $E_N(\rho)$ was defined in \cref{eq:logneg-defn}. Note that $\tilde{E}_1 \equiv E_N(\bar{\tilde{\rho}}^R_\alpha) = E_1$ by definition.}
\label{tab:neg}
\end{table*}

\begin{figure*}[htbp]
\centering
\includegraphics[width=0.48\linewidth]{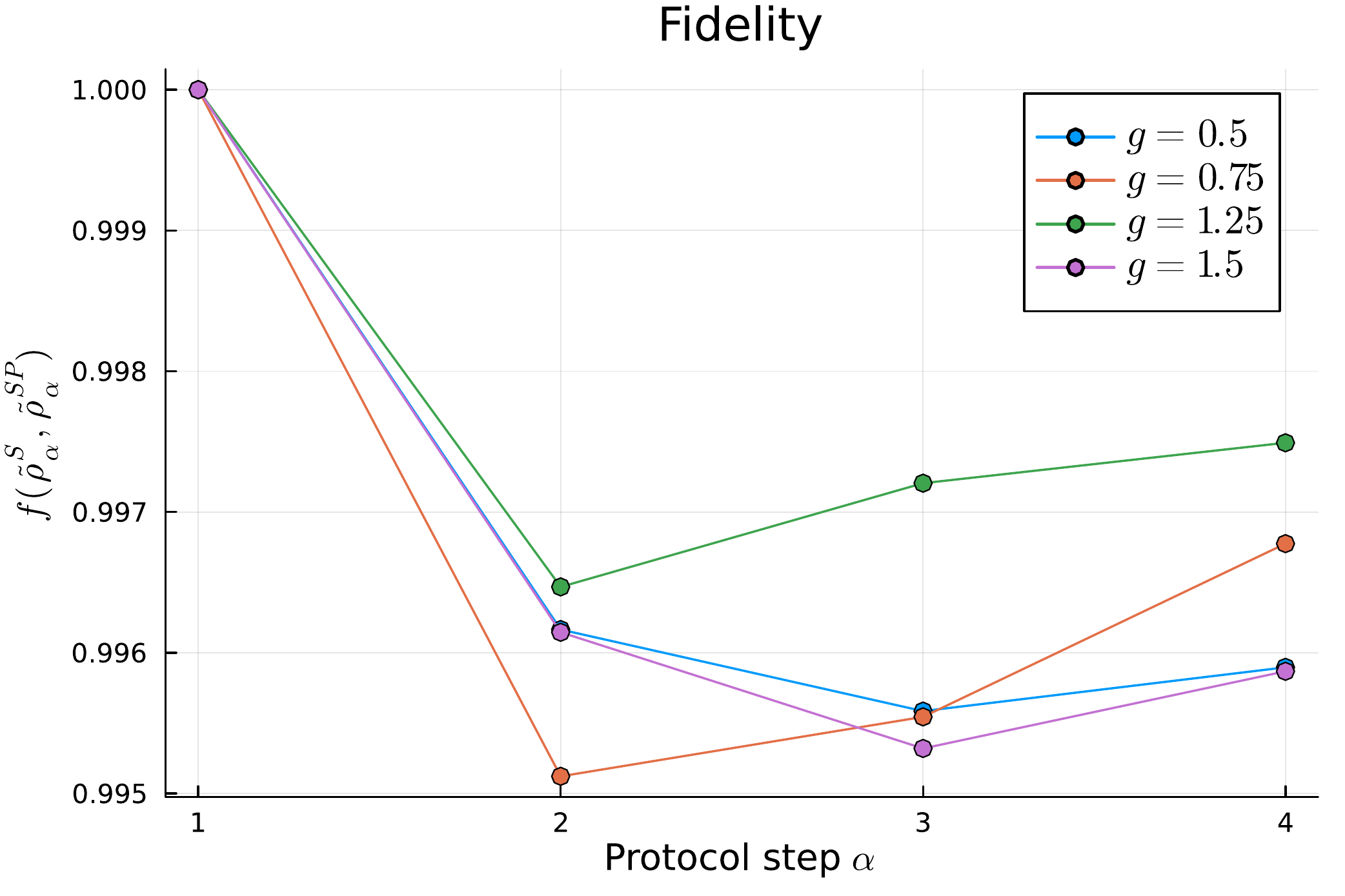} 
\includegraphics[width=0.48\linewidth]{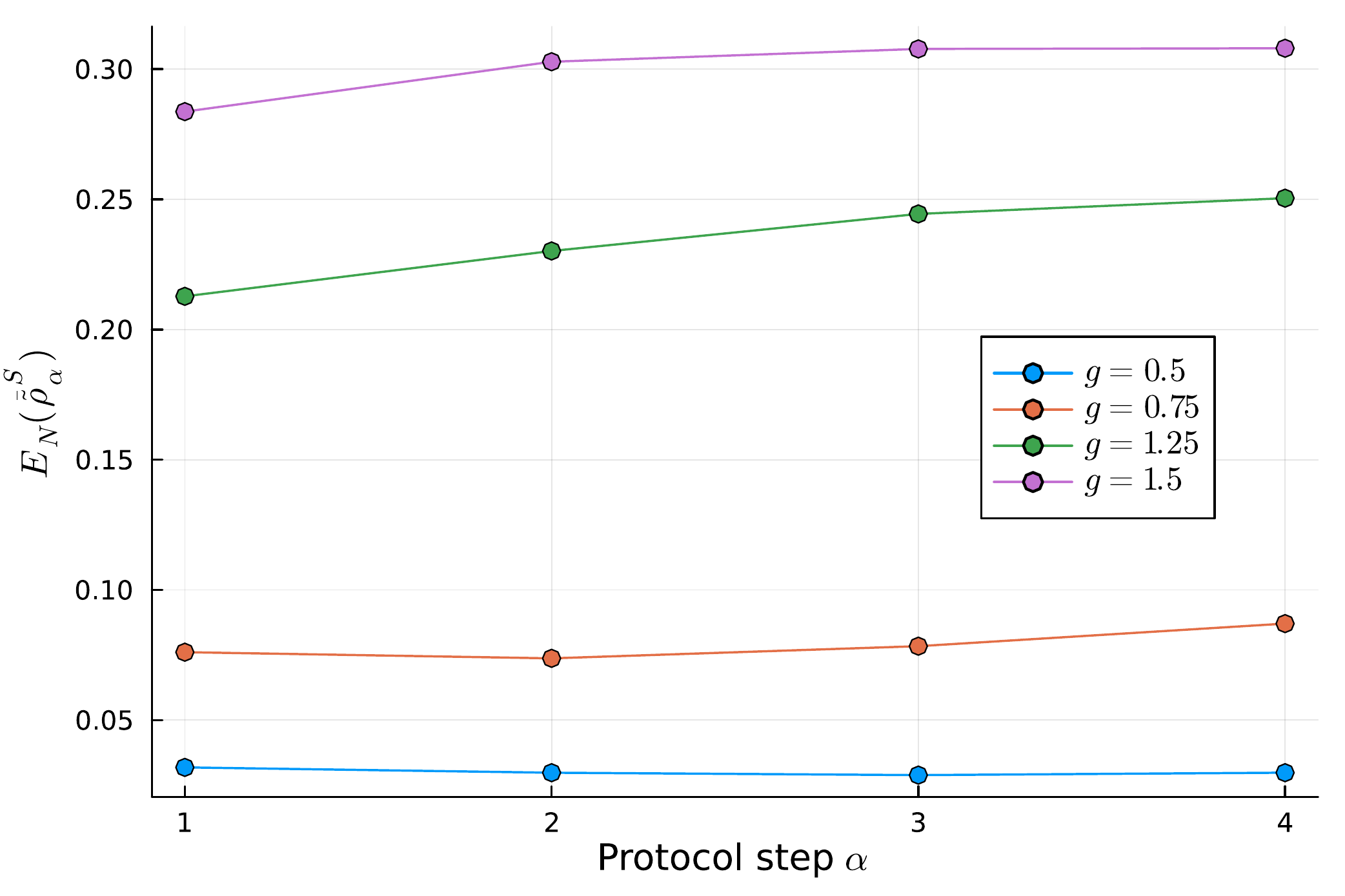}
\caption{\label{fig:fid+neg} The plot the fidelity per spin-pair $f(\tilde{\rho}^S_\alpha,\tilde{\rho}^{SP}_\alpha)$ (left), and the negativity $E_N(\bar{\tilde{\rho}}^S_\alpha)$ as a function of the number of spin-pairs extracted.  Left panel: Data is given in \cref{tab:fid2} under $\tilde f_{S, SP}$. This figure shows that the symmetrized state is close to a product state.  Right panel: $\tilde E_3$ in \cref{tab:neg}.}
\end{figure*}

\subsection{Results: Multiple steps}
\label{sec:prot-multi}

Having investigated the rise of entanglement in the register after the first swap, we now construct a protocol to build up entanglement in the registers as a function of $\alpha$. Since ${\cal N}_1(t)$, defined in the previous section, is maximized when $t = t_{\rm min}(g)$, we choose to perform the $\alpha$th swap at the times $t_\alpha = \alpha t_{\rm min}(g)$ for each value of $g$. We can then compute both the density matrices $\rho^R_\alpha$ and $\tilde{\rho}^R_\alpha$ to study how the entanglement depends on $\alpha$.

Any protocol to extract entanglement from the registers will need access to the density matrix $\rho^R_\alpha$. Unfortunately, there is no practical way to measure the density matrix of a single quantum system without destroying the state. Measurement of density matrices are, however, possible with access to an infinite number of independent identical systems. In fact to use the distillation protocol defined in \cref{eq:distillation}, we already assume the quantum state of the register to be in a direct product of independent and identical quantum systems, each of size $b$ spin-pairs. This means we would ideally like to have
\begin{align}
\rho^R_\alpha =\ \underbrace{\bar{\rho}\otimes \bar{\rho} \otimes ... \otimes \bar{\rho}}_{(\alpha+1) \text{ copies}},
\label{eq:rhoR}
\end{align}
where $\bar{\rho}$ is a $2b\times 2b$ matrix whose properties can be used to practically to extract entanglement especially when $\alpha$ is large. This form implies two results: first, all the \(\alpha+1\) \(b\) spin pairs are in an identical state, and second, they are independent so that \(\rho_\alpha^R\) is a direct product. To check the first, we compute $\bar{\rho}^R_i,i=0,1,2,..,\alpha$ as the $2^{2b}\times 2^{2b}$ matrix of the $i$th $b$ spin-pair obtained by tracing $\rho^R_\alpha$ over the remaining spin-pairs and check how similar $\bar{\rho}^R_i,i=0,1,2,..,\alpha$ are to each other. One way to check this is to compute the fidelity $F(\bar{\rho}^R_i,\bar{\rho}^R_j)$ between these matrices where the Fidelity is defined through the expression\footnote{This is the quantum analog of the square of the classical Bhattacharya coefficient~\cite{Bhattacharyya:1943} that provides~\cite{Matusita:1971} an upper bound to the total probability of misclassification in a symmetric Bayesian context.}
\begin{align}
F(\rho_i,\rho_j) \ & = \ \Big(\mathrm{tr} \Big[\sqrt{\sqrt{\rho_i}\rho_j \sqrt{\rho}_i}\Big]\Big)^2.
\label{eq:fid}
\end{align}
If the two density matrices are similar, then this quantity will be close to one. Second we need to check if $\rho^R_\alpha$ is indeed the same as density matrix constructed as 
\begin{align}
\rho^P_\alpha =\ \bar{\rho}^R_0 \otimes \bar{\rho}^R_1 ... \otimes \bar{\rho}^R_\alpha.
\label{eq:rhoP}
\end{align}
We can again do this by computing the Fidelity $F(\rho^R_\alpha,\rho^P_\alpha)$ and making sure that it is close to one. 

Finally, if we do not distinguish between the \(\alpha+1\) \(b\) spin pairs in our distillation protocol, we can choose to consider a symmetrized density matrix $\rho^S_\alpha$ as discussed in \cref{app:perm}. Due to the quantum de Finniti theorem~\cite{hudson1976locally,QdF}, this symmetrized density matrix is known to take the form 
\begin{align}
\rho^S_\alpha\ =\ \underbrace{\bar{\rho} \otimes \bar{\rho} \otimes ... \otimes \bar{\rho}}_{(\alpha+1) \text{ copies}},
\label{eq:rhos}
\end{align}
in the limit $\alpha\rightarrow \infty$. Here $\bar{\rho}$ is a $2^{2b} \times 2^{2b}$ density matrix of $b$ spin-pairs, while $\rho^R_\alpha$ was a $2^{2b(\alpha+1)} \times 2^{2b(\alpha+1)}$ matrix. When $\alpha$ is not infinite we do not expect $\rho^S_\alpha$ to be of the form \cref{eq:rhos} but we can again ask how close to it do we get. As in the case of $\rho^R_\alpha$, we can again compute $\bar{\rho}^S_i,i=0,1,...,\alpha$. But since $\rho^S_\alpha$ is symmetrized we will find $\bar{\rho}^S_0=\bar{\rho}^S_1=...=\bar{\rho}^S_\alpha$. We can also construct
\begin{align}
\rho^{SP}_\alpha =\ \underbrace{\bar{\rho}^S_\alpha \otimes \bar{\rho}^S_\alpha ... \otimes \bar{\rho}^S_\alpha}_{(\alpha+1) \text{ copies}}
\label{eq:rhosp}
\end{align}
and compare it with $\rho^S_\alpha$ for various $\alpha$'s. Note that we defined $\rho^P_\alpha$, $\rho^S_\alpha$, $\rho^{SP}_\alpha$ starting from $\rho^R_\alpha$. We can similarly define
$\tilde{\rho}^P_\alpha$, $\tilde{\rho}^S_\alpha$, $\tilde{\rho}^{SP}_\alpha$ starting from $\tilde{\rho}^R_\alpha,\alpha = 1,2,..$ which is obtained from $\rho^R_\alpha$ by tracing over the zeroth $b$ spin-pair (i.e., $\tilde{\rho}^R_0=1$). Note that $\tilde{\rho}^R_\alpha$ will be a $2^{2b\alpha}\times 2^{2b\alpha}$ matrix. We will see below that $\tilde{\rho}^R_\alpha$ and $\tilde{\rho}^S_\alpha$ are closer to the form described by \cref{eq:rhoR} and \cref{eq:rhos} with $\alpha$-copies than $\rho^R_\alpha$ and $\rho^S_\alpha$.

In our work we choose $b=1$, $\delta=1$, $L=16$ and assume we have up to $N=5$ spin-pairs in the register. Thus we can compute the sequence of density matrices $\rho^R_{\alpha},\alpha=0,1,2,3,4$ of sizes $4^{(\alpha+1)}\times 4^{(\alpha+1)}$. For $\alpha=4$ we can compute $\bar{\rho}^R_i,i=0,1,2,3,4$ which are $4\times 4$ matrices. In \cref{tab:fid1} we show the results for $F(\bar{\rho}_i,\bar{\rho}_j)$ at $g=0.5,1.0$ and $1.5$. Based on these results we note that $\bar{\rho}^R_0$, obtained after the first swap, is quite different from the remaining. The other four density matrices are very similar. This suggests that it may be better to trace over $\alpha=0$ spin-pairs and focus on $\tilde{\rho}^R_\alpha$. 

\begin{table*}[htbp]
\renewcommand{\arraystretch}{1.0}
\input{tables/tab4}
\caption{The maximum singlet fraction \(F_m\) of $i$-th extracted spin pair computed using $\bar{\rho}^R_i$. In each case we also give the angles to the closest maximally entangled state, parameterized through \cref{eq:BellParameters}.}
\label{tab:singlet}
\end{table*}

\begin{table*}[htbp]
\renewcommand{\arraystretch}{1.0}
\input{tables/tab5}
\caption{The maximum singlet fraction \(F_m\) computed using $\bar{\rho}^S_\alpha$ and the closest maximally entangled state, parameterized as in \cref{eq:BellParameters}. Similarly \(\tilde{F}_m\) refers to the maximal singlet fraction computed using $\bar{\tilde{\rho}}^S_\alpha$ where the zeroth spin-pair was ignored. Note that all unentangled states have \(f\leq0.5\)~\cite{PhysRevA.66.022307}, the actual entanglement negativities are presented in \cref{tab:neg}.}
\label{tab:avsinglet}
\end{table*}

In \cref{tab:fid2} we compare the matrices $\rho^R_\alpha$, $\rho^P_\alpha$, $\rho^S_\alpha$ and $\rho^{SP}_\alpha$. Since these are density matrices involving different number of spin-pairs, instead of using the fidelity defined in \cref{eq:fid}, we define the fidelity per spin-pair using 
the $k$th root of the Fidelity defined as
\begin{align}
f(\rho,\rho') = (F(\rho,\rho'))^{1/k}
\label{eq:fidk}
\end{align} 
where $k$ labels the number of spin-pairs that the density matrices $\rho$ and $\rho'$ describe. We notice again that dropping the zeroth spin-pair is very helpful and $\tilde{\rho}^R_\alpha$ is well described by a product state and symmetrization helps improve things. In \cref{fig:fid+neg} we plot the fidelity $f(\tilde{\rho}^S_\alpha,\tilde{\rho}^{SP}_\alpha)$ and the logarithmic negativity $E_N(\bar{\tilde{\rho}}^S_\alpha)$ as a function of $\alpha$.

As described in \cref{app:perm}, most distillation protocols only have access to the symmetrized part of the density matrix of spin-pairs extracted at different times. While the different spin-pairs can be entangled to roughly the same extent, they can be aligned in different directions. In such situations the symmetrized density matrix is likely to be less entangled and can even be unentangled. This can be seen in our results in \cref{tab:neg}, where we compute the logarithmic negativity extracted from $\bar{\rho}^R_\alpha$, $\rho^R_\alpha$ and $\bar{\rho}^S_\alpha$. Note that while $E_N(\bar{\rho}^R_1)$ and $E_N(\rho^R_1)/2$ are non-zero, $E_N(\bar{\rho}^S_1)$ vanishes. This means that though the zeroth spin-pair and the first spin-pair are both entangled, symmetrization kills this entanglement, suggesting that both spin-pairs are entangled but aligned differently. We discuss this issue of alignment more clearly in \cref{app:align} by defining the maximal singlet fraction $F_m(\rho)$ for a given density matrix $\rho$ and explain how we can use it to construct the density matrix $\rho^A$ that is aligned properly. So instead of symmetrizing $\rho^R_\alpha$ it may actually be necessary to symmetrize $\rho^A_\alpha$ while constructing $\rho^{S}_\alpha$. Let us explore if this is necessary in our case. 

In order to study how the various extracted spin-pairs are aligned, in \cref{tab:singlet} we compute the singlet fraction $F_m$ of each extracted spin-pair using $\bar{\rho}^R_i,i=0,1,2,3,4$ obtained from $\rho^R_{\alpha=4}$. We also compute the nearest corresponding maximal Bell-state using three angles $\theta$, $\phi$ and $\beta$ as defined in \cref{eq:BellParameters}. Note that although $F_m$ is roughly the same for all the five spin-pairs, the closest maximally entangled state for the zeroth pair is aligned very differently than the remaining pairs which are more closely aligned. This is the reason that in \cref{tab:neg}, while $E_N(\bar{\rho}^R_0)$ and $E_N(\bar{\rho}^R_1)$ are non-zero showing these spin-pairs are entangled, $E_N(\bar{\rho}^S_1)$ vanished since symmetrization killed the negativity. This is again consistent with our previous observation that it is best to discard the zeroth spin-pair. The importance of dropping the zeroth spin-pair is also visible in \cref{tab:avsinglet}, where we compute the maximal singlet fraction $F_m$ for each spin pair obtained from the symmetrized density matrix, $\bar{\rho}^S_\alpha$ and compare it with $\bar{\tilde{\rho}}^S_\alpha$ where the zeroth spin pair is dropped. When we keep the zeroth spin-pair during the symmetrization procedure, $F_m$ initially drops and then slowly begins to build back as more and more aligned spin-pairs are added to the register. On the other hand if we ignore the zeroth spin-pair, $F_m$ changes only slightly as the spin-pairs accumulate in the register.  This suggests that we do not need to explicitly align the spin pairs in our protocol if we ignore the zeroth spin pair.


%% file: tables/tab_fidelities.tex
\begin{tabular}{lllll}
\TopRule
\multicolumn{5}{c}{ $ g=0.5, L=16 $ }\\
\MidRule
    1.000 & 0.708 & 0.710 & 0.713 & 0.715 \\
      & 1.000 & 0.999 & 0.997 & 0.994 \\
      &       & 1.000 & 0.999 & 0.997 \\
      &       &       & 1.000 & 0.999 \\
      &       &       &       & 1.000 \\
\BotRule
\end{tabular}
\hspace{0.2em}
\begin{tabular}{lllll}
\TopRule
\multicolumn{5}{c}{ $ g=0.75, L=16 $ }\\
\MidRule
    1.000 & 0.710 & 0.720 & 0.727 & 0.728 \\
      & 1.000 & 0.997 & 0.990 & 0.981 \\
      &       & 1.000 & 0.997 & 0.991 \\
      &       &       & 1.000 & 0.998 \\
      &       &       &       & 1.000 \\
\BotRule
\end{tabular}
\hspace{0.2em}
\begin{tabular}{lllll}
\TopRule
\multicolumn{5}{c}{ $ g=1.25, L=16  $ }\\
\MidRule
    1.000 & 0.700 & 0.760 & 0.768 & 0.760 \\
      & 1.000 & 0.988 & 0.984 & 0.988 \\
      &       & 1.000 & 1.000 & 1.000 \\
      &       &       & 1.000 & 1.000 \\
      &       &       &       & 1.000 \\
\BotRule
\end{tabular}
\hspace{0.2em}
\begin{tabular}{lllll}
\TopRule
\multicolumn{5}{c}{ $ g=1.5, L=16 $ }\\
\MidRule
    1.000 & 0.700 & 0.766 & 0.751 & 0.741 \\
      & 1.000 & 0.990 & 0.994 & 0.996 \\
      &       & 1.000 & 0.999 & 0.998 \\
      &       &       & 1.000 & 1.000 \\
      &       &       &       & 1.000 \\
\BotRule
\end{tabular}

%% file: tables/tab2.tex
\begin{tabular}{c|ccc|ccc}
\TopRule
\multicolumn{7}{c}{ $ g = 0.50, \ L = 16 $ }\\
\MidRule
 \(\alpha\)&\(f_{R,S}\)&\(f_{R,P}\)&\(f_{S,SP}\)&\(\tilde f_{R,S}\)&\(\tilde f_{R,P}\)&\(\tilde f_{S,SP}\)\\
\MidRule
0 & 1.000 & 1.000 & 1.000 & - & - & -  \\
1 & 0.923 & 0.991 & 0.969 & 1.000 & 1.000 & 1.000 \\
2 & 0.922 & 0.989 & 0.974 & 1.000 & 0.996 & 0.996 \\
3 & 0.929 & 0.988 & 0.978 & 0.999 & 0.994 & 0.996 \\
4 & 0.935 & 0.989 & 0.982 & 0.998 & 0.994 & 0.996 \\
\BotRule
\end{tabular}
\begin{tabular}{c|ccc|ccc}
\TopRule
\multicolumn{7}{c}{ $ g = 0.75, \ L = 16 $ }\\
\MidRule
 \(\alpha\)&\(f_{R,S}\)&\(f_{R,P}\)&\(f_{S,SP}\)&\(\tilde f_{R,S}\)&\(\tilde f_{R,P}\)&\(\tilde f_{S,SP}\)\\
\MidRule
0 & 1.000 & 1.000 & 1.000 & - & - & -  \\
1 & 0.925 & 0.986 & 0.960 & 1.000 & 1.000 & 1.000 \\
2 & 0.923 & 0.983 & 0.970 & 0.999 & 0.994 & 0.995 \\
3 & 0.928 & 0.983 & 0.978 & 0.996 & 0.992 & 0.996 \\
4 & 0.933 & 0.985 & 0.984 & 0.993 & 0.992 & 0.997 \\
\BotRule
\end{tabular}
\vskip0.1in
\begin{tabular}{c|ccc|ccc}
\TopRule
\multicolumn{7}{c}{ $ g = 1.25, \ L = 16 $ }\\
\MidRule
 \(\alpha\)&\(f_{R,S}\)&\(f_{R,P}\)&\(f_{S,SP}\)&\(\tilde f_{R,S}\)&\(\tilde f_{R,P}\)&\(\tilde f_{S,SP}\)\\
\MidRule
0 & 1.000 & 1.000 & 1.000 & - & - & -  \\
1 & 0.926 & 0.983 & 0.949 & 1.000 & 1.000 & 1.000 \\
2 & 0.920 & 0.979 & 0.971 & 0.991 & 0.991 & 0.996 \\
3 & 0.922 & 0.978 & 0.983 & 0.986 & 0.988 & 0.997 \\
4 & 0.926 & 0.977 & 0.988 & 0.983 & 0.985 & 0.997 \\
\BotRule
\end{tabular}
\begin{tabular}{c|ccc|ccc}
\TopRule
\multicolumn{7}{c}{ $ g = 1.50, \ L = 16 $ }\\
\MidRule
 \(\alpha\)&\(f_{R,S}\)&\(f_{R,P}\)&\(f_{S,SP}\)&\(\tilde f_{R,S}\)&\(\tilde f_{R,P}\)&\(\tilde f_{S,SP}\)\\
\MidRule
0 & 1.000 & 1.000 & 1.000 & - & - & -  \\
1 & 0.924 & 0.985 & 0.950 & 1.000 & 1.000 & 1.000 \\
2 & 0.916 & 0.977 & 0.973 & 0.987 & 0.987 & 0.996 \\
3 & 0.915 & 0.973 & 0.983 & 0.982 & 0.982 & 0.995 \\
4 & 0.917 & 0.970 & 0.988 & 0.978 & 0.977 & 0.996 \\
\BotRule
\end{tabular}

%% file: tables/tab3.tex
\begin{tabular}{c|ccc|cc}
\TopRule
\multicolumn{6}{c}{ $ g=0.5,\ L=16 $ }\\
\MidRule
$\alpha$ & $E_1$ & $E_2$ & $E_3$ & $\tilde E_2$ & $\tilde E_3$ \\
\MidRule
     0 &   0.05 &   0.05 &   0.05 &      - &      - \\
     1 &   0.03 &   0.07 &   0.00 &   0.03 &   0.03 \\ 
     2 &   0.03 &   0.07 &   0.00 &   0.04 &   0.03 \\ 
     3 &   0.03 &   0.07 &   0.00 &   0.04 &   0.03 \\ 
     4 &   0.05 &   0.07 &   0.01 &   0.05 &   0.03 \\ 
\BotRule
\end{tabular}
\hspace{1em}
\begin{tabular}{c|ccc|cc}
\TopRule
\multicolumn{6}{c}{ $ g=0.75,\ L=16 $ }\\
\MidRule
$\alpha$ & $E_1$ & $E_2$ & $E_3$ & $\tilde E_2$ & $\tilde E_3$ \\
\MidRule
     0 &   0.12 &   0.12 &   0.12 &      - &      - \\
     1 &   0.08 &   0.14 &   0.00 &   0.08 &   0.08 \\ 
     2 &   0.08 &   0.14 &  -0.00 &   0.09 &   0.07 \\ 
     3 &   0.11 &   0.14 &   0.00 &   0.11 &   0.08 \\ 
     4 &   0.15 &   0.14 &   0.01 &   0.12 &   0.09 \\ 
\BotRule
\end{tabular}
\hspace{1em}
\begin{tabular}{c|ccc|cc}
\TopRule
\multicolumn{6}{c}{ $ g=1.25,\ L=16  $ }\\
\MidRule
$\alpha$ & $E_1$ & $E_2$ & $E_3$ & $\tilde E_2$ & $\tilde E_3$ \\
\MidRule
     0 &   0.33 &   0.33 &   0.33 &      - &      - \\
     1 &   0.21 &   0.29 &   0.00 &   0.21 &   0.21 \\ 
     2 &   0.27 &   0.30 &  -0.00 &   0.26 &   0.23 \\ 
     3 &   0.29 &   0.31 &   0.02 &   0.28 &   0.24 \\ 
     4 &   0.27 &   0.31 &   0.06 &   0.28 &   0.25 \\ 
\BotRule
\end{tabular}
\hspace{1em}
\begin{tabular}{c|ccc|cc}
\TopRule
\multicolumn{6}{c}{ $ g=1.5,\ L=16 $ }\\
\MidRule
$\alpha$ & $E_1$ & $E_2$ & $E_3$ & $\tilde E_2$ & $\tilde E_3$ \\
\MidRule
     0 &   0.32 &   0.32 &   0.32 &      - &      - \\
     1 &   0.28 &   0.32 &  -0.00 &   0.28 &   0.28 \\ 
     2 &   0.34 &   0.34 &   0.01 &   0.33 &   0.30 \\ 
     3 &   0.32 &   0.35 &   0.07 &   0.34 &   0.31 \\ 
     4 &   0.31 &   0.35 &   0.11 &   0.35 &   0.31 \\ 
\BotRule
\end{tabular}

%% file: tables/tab4.tex
\begin{tabular}{rrrrr}
\TopRule
\multicolumn{5}{c}{\(g = 0.5,\ L = 16\)}\\
\MidRule
\(i\)&\(F_m\)&\(\theta\)&\(\phi\)&\(\beta\)\\
\MidRule
0&\(0.52\)&\(0.79\)&\(0.00\)&\(1.57\)\\
1&\(0.51\)&\(0.72\)&\(0.34\)&\(-0.60\)\\
2&\(0.51\)&\(0.73\)&\(0.32\)&\(-0.43\)\\
3&\(0.51\)&\(0.74\)&\(0.30\)&\(-0.19\)\\
4&\(0.52\)&\(0.75\)&\(0.28\)&\(0.00\)\\
5&\(0.52\)&\(0.75\)&\(0.27\)&\(0.13\)\\
6&\(0.53\)&\(0.75\)&\(0.27\)&\(0.20\)\\
\BotRule
\end{tabular}
\hspace{0.2em}
\begin{tabular}{rrrrr}
\TopRule
\multicolumn{5}{c}{\(g = 0.75,\ L = 16\)}\\
\MidRule
\(i\)&\(F_m\)&\(\theta\)&\(\phi\)&\(\beta\)\\
\MidRule
0&\(0.54\)&\(0.79\)&\(0.00\)&\(1.57\)\\
1&\(0.52\)&\(0.66\)&\(0.46\)&\(-0.44\)\\
2&\(0.53\)&\(0.69\)&\(0.42\)&\(-0.14\)\\
3&\(0.54\)&\(0.71\)&\(0.38\)&\(0.11\)\\
4&\(0.55\)&\(0.71\)&\(0.37\)&\(0.24\)\\
5&\(0.56\)&\(0.71\)&\(0.36\)&\(0.28\)\\
\BotRule
\end{tabular}
\hspace{0.2em}
\begin{tabular}{rrrrr}
\TopRule
\multicolumn{5}{c}{\(g = 1.25,\ L = 16\)}\\
\MidRule
\(i\)&\(F_m\)&\(\theta\)&\(\phi\)&\(\beta\)\\
\MidRule
0&\(0.26\)&\(0.00\)&\(0.79\)&\(0.01\)\\
1&\(0.58\)&\(0.60\)&\(0.54\)&\(0.07\)\\
2&\(0.60\)&\(0.65\)&\(0.48\)&\(0.36\)\\
3&\(0.61\)&\(0.65\)&\(0.47\)&\(0.39\)\\
4&\(0.60\)&\(0.65\)&\(0.48\)&\(0.36\)\\
\BotRule
\end{tabular}
\hspace{0.2em}
\begin{tabular}{rrrrr}
\TopRule
\multicolumn{5}{c}{\(g = 1.5,\ L = 16\)}\\
\MidRule
\(i\)&\(F_m\)&\(\theta\)&\(\phi\)&\(\beta\)\\
\MidRule
0&\(0.62\)&\(0.79\)&\(0.00\)&\(-1.57\)\\
1&\(0.61\)&\(0.59\)&\(0.55\)&\(0.28\)\\
2&\(0.63\)&\(0.64\)&\(0.49\)&\(0.49\)\\
3&\(0.62\)&\(0.63\)&\(0.50\)&\(0.44\)\\
4&\(0.62\)&\(0.63\)&\(0.51\)&\(0.41\)\\
\BotRule
\end{tabular}

%% file: tables/tab5.tex
\begin{tabular}{r|rrrr|rrrr}
\TopRule
\multicolumn{8}{c}{\(g = 0.5\), \(L = 16\)}\\
\MidRule
&\multicolumn{4}{c|}{keeping zeroth spin}&\multicolumn{4}{c}{dropping zeroth spin}\\
\MidRule
\(\alpha\)&\(F_m\)&\(\theta\)&\(\phi\)&\(\beta\)&\(F_m\)&\(\theta\)&\(\phi\)&\(\beta\)\\
\MidRule
0&\(0.52\)&\(0.79\)&\(0.00\)&\(1.57\)&-&-&-&-\\
1&\(0.48\)&\(0.77\)&\(0.19\)&\(-1.43\)&\(0.51\)&\(0.72\)&\(0.34\)&\(-0.60\)\\
2&\(0.48\)&\(0.75\)&\(0.26\)&\(-0.81\)&\(0.51\)&\(0.72\)&\(0.33\)&\(-0.52\)\\
3&\(0.49\)&\(0.75\)&\(0.26\)&\(-0.43\)&\(0.51\)&\(0.73\)&\(0.32\)&\(-0.41\)\\
4&\(0.49\)&\(0.75\)&\(0.25\)&\(-0.24\)&\(0.51\)&\(0.73\)&\(0.31\)&\(-0.30\)\\
\BotRule
\end{tabular}
\hspace{0.2em}
\begin{tabular}{r|rrrr|rrrr}
\TopRule
\multicolumn{8}{c}{\(g = 0.75\), \(L = 16\)}\\
\MidRule
&\multicolumn{4}{c|}{keeping zeroth spin}&\multicolumn{4}{c}{dropping zeroth spin}\\
\MidRule
\(\alpha\)&\(F_m\)&\(\theta\)&\(\phi\)&\(\beta\)&\(F_m\)&\(\theta\)&\(\phi\)&\(\beta\)\\
\MidRule
0&\(0.54\)&\(0.79\)&\(0.00\)&\(1.57\)&-&-&-&-\\
1&\(0.47\)&\(0.75\)&\(0.26\)&\(-1.51\)&\(0.52\)&\(0.66\)&\(0.46\)&\(-0.44\)\\
2&\(0.46\)&\(0.70\)&\(0.39\)&\(-0.62\)&\(0.52\)&\(0.67\)&\(0.44\)&\(-0.29\)\\
3&\(0.48\)&\(0.72\)&\(0.34\)&\(-0.05\)&\(0.53\)&\(0.68\)&\(0.42\)&\(-0.15\)\\
4&\(0.49\)&\(0.73\)&\(0.33\)&\(0.09\)&\(0.53\)&\(0.69\)&\(0.41\)&\(-0.04\)\\
\BotRule
\end{tabular}
\hspace{0.2em}
\begin{tabular}{r|rrrr|rrrr}
\TopRule
\multicolumn{8}{c}{\(g = 1.25\), \(L = 16\)}\\
\MidRule
&\multicolumn{4}{c|}{keeping zeroth spin}&\multicolumn{4}{c}{dropping zeroth spin}\\
\MidRule
\(\alpha\)&\(F_m\)&\(\theta\)&\(\phi\)&\(\beta\)&\(F_m\)&\(\theta\)&\(\phi\)&\(\beta\)\\
\MidRule
0&\(0.26\)&\(0.00\)&\(0.79\)&\(0.01\)&-&-&-&-\\
1&\(0.43\)&\(0.75\)&\(0.26\)&\(-1.72\)&\(0.58\)&\(0.60\)&\(0.54\)&\(0.07\)\\
2&\(0.45\)&\(0.64\)&\(0.50\)&\(0.49\)&\(0.59\)&\(0.62\)&\(0.52\)&\(0.22\)\\
3&\(0.49\)&\(0.64\)&\(0.49\)&\(0.44\)&\(0.59\)&\(0.63\)&\(0.50\)&\(0.28\)\\
4&\(0.51\)&\(0.64\)&\(0.49\)&\(0.41\)&\(0.59\)&\(0.64\)&\(0.50\)&\(0.30\)\\
\BotRule
\end{tabular}
\hspace{0.2em}
\begin{tabular}{r|rrrr|rrrr}
\TopRule
\multicolumn{8}{c}{\(g = 1.5\), \(L = 16\)}\\
\MidRule
&\multicolumn{4}{c|}{keeping zeroth spin}&\multicolumn{4}{c}{dropping zeroth spin}\\
\MidRule
\(\alpha\)&\(F_m\)&\(\theta\)&\(\phi\)&\(\beta\)&\(F_m\)&\(\theta\)&\(\phi\)&\(\beta\)\\
\MidRule
0&\(0.62\)&\(0.79\)&\(0.00\)&\(1.57\)&-&-&-&-\\
1&\(0.43\)&\(0.66\)&\(0.46\)&\(0.96\)&\(0.61\)&\(0.59\)&\(0.55\)&\(0.28\)\\
2&\(0.49\)&\(0.61\)&\(0.53\)&\(0.65\)&\(0.62\)&\(0.61\)&\(0.52\)&\(0.39\)\\
3&\(0.52\)&\(0.61\)&\(0.53\)&\(0.56\)&\(0.62\)&\(0.62\)&\(0.52\)&\(0.41\)\\
4&\(0.54\)&\(0.61\)&\(0.53\)&\(0.52\)&\(0.62\)&\(0.62\)&\(0.52\)&\(0.41\)\\
\BotRule
\end{tabular}

%% file: sec_conclusions.tex
\section{Conclusions}
\label{sec:conclusions}
We have studied the dynamics of entanglement in the ground state of the one-dimensional \ac{TFIM}, in particular, how the entanglement recovers after it is externally set to zero. We find that the entanglement rebounds and oscillates, asymptotically reaching its ground-state value at long times. This allows us to repeatedly extract this entanglement into a register---and the extracted entanglement can be optimized by choosing the interval between extractions to be equal to the time of the first rebound peak.  Finally, we find that the pairs extracted near this peak of each rebound are almost aligned, and can be used in a standard protocol to distill Bell pairs.

In this first study, we could only look at single spin-pairs separated by a distance of one lattice spacing in a one dimensional spin chain.  Further work is necessary to see how the results generalize to the interesting case of  blocks of spins-pairs and separations that are a non-zero fraction of the correlation length. We can then connect the results with entanglement extraction from continuum quantum field theory that emerges as one approaches the critical point at $g=1$.
Numerically, this requires both large block sizes $b$ and separations $\delta$ with $\delta/b$ kept fixed.  

An important feature of this approach to entanglement harvesting protocol is 
its feasibility on near-term analog and digital quantum devices. For example, Ising-like interactions can be natively generated on cold-atom simulators based on Rydberg atoms \cite{schauss_quantum_2018}, and local entangling operations between qubits have been demonstrated \cite{semeghini_probing_2021}. It would be interesting to explore whether this simple protocol can be implemented on quantum hardware to generate Bell pairs from the vacuum of a spin chain.

\section*{Acknowledgements}

HS would like to thank the developers of ITensor library \cite{itensor-r0.3, itensor} for their excellent work, and especially Matthew Fishman for very useful conversations regarding tensor networks.
HS would also like to thank Anthony Ciavarella, Natalie Klco and Martin Savage for discussions on entanglement negativity.
We would like to thank Alex Buser for early collaboration on this work. 
HS is supported in part by the DOE QuantISED program through the theory consortium ``Intersections of QIS and Theoretical Particle Physics'' at Fermilab with Fermilab Subcontract No. 666484,
in part by the Institute for Nuclear Theory with US Department of Energy Grant DE-FG02-00ER41132,
and in part by the U.S. Department of Energy, Office of Science, Office of Nuclear Physics, Inqubator for Quantum Simulation (IQuS) under Award Number DOE (NP) Award DE-SC0020970.
The material presented here is based on work supported by the U.S. Department of Energy, Office of Science---High Energy Physics Contract KA2401032 (Triad National Security, LLC Contract Grant No. 89233218CNA000001) to Los Alamos National Laboratory.
S.C. is supported by a Duke subcontract of this grant. S.C is also supported in part by the U.S. Department of Energy, Office of Science, Nuclear Physics program under Award No. DE-FG02-05ER41368.


%% file: app_effdens.tex
\section{Density Matrices}
\label{app:effdens}

In this section we review some well known ideas about density matrices, to help a reader who may not be familiar with them. First we note that it is always possible to reformulate everything in quantum mechanics using the concept of a density matrix $\rho$ of a quantum state. For example, given a normalized state $|\psi\rangle$ in the Hilbert space of a quantum system, we can construct the density matrix \(\rho \equiv |\psi\rangle\langle \psi|\). It is then easy to see that the operator $|\psi\rangle  \to \hat{O} |\psi\rangle$ can be implemented directly on the density matrix as $\rho \to \hat{O} \rho \hat{O}^\dagger$.  Moreover, the probability that observable $\hat{O}$ takes the value $o$ in a projective measurement is given by 
\begin{align}
\sum_{\psi^o_i}|\langle\psi|\psi^o_i \rangle|^2 \equiv \Tr \rho P^o\,,
\end{align}
where \(\{|\psi^o_i\rangle\}\) is a basis for the eigen-subspace of \(\hat O\) with eigenvalue \(o\) and $P^o \equiv \sum_{\psi^o_i}|\psi^o_i\rangle\langle\psi^o_i |$ is the projector into this eigen-subspace\footnotepunct{The entire description can be generalized to non-projective measurements using positive operator-valued measures~\cite{doi:10.1119/1.19280}}. In addition, in this case, the posterior state conditioned on the observation is given by \(\rho P^o/\Tr \rho P^o\). In other words, all properties of a pure state in the Hilbert space of a quantum system can be studied using the density matrix.

When we are dealing with subsystems of quantum systems, it is more economical to introduce the concept of a density matrix, since pure states would require us to consider the entire quantum system. If we are only interested in the 
subsystem, which could be much smaller than the entire quantum system, then we can treat the subsystem as a quantum system in a mixed state described by a density matrix. We can define the density matrix of the entire system as the full density matrix, while the density matrix of the subsystem as the reduced density matrix obtained as the partial trace of the full density matrix over the Hilbert space complementary to the subsystem.

The equivalence of the two pictures---dealing with the density matrix of the entire subsystem or the reduceded density matrix of only the subsystem of interest---is most easily explained by the notion of an effective density matrix, which also generalizes the concept of the reduced density matrix. Here we explain this idea, as motivated in \citet[Section 3.2]{Casini:2013rba}, but generalized to be more appropriate for our work. The idea of an effective density matrix arises from the fact that a subsystem is associated with a set of observables that act only on it, and thus with an algebra of a subset of operators \(\{\hat O\}\) of interest, all of which are invariant under an unitary group of transformations \(\{U\}\) of the complementary subsystems:
\begin{equation}
   U \hat O U^\dagger = \hat O \qquad 
\end{equation}
for all $U$ and  $\hat O$. Then, for any density matrix \(\rho\),
\begin{align}
  \Tr \hat O \rho &= \Big\langle \Tr \big( U\ \hat O \ U^\dagger  \rho \big) \Big\rangle_U\ =\  \Tr \hat O \rho^e\,,
  \label{eq:effective}
\end{align}
where $\langle \ \dots\ \rangle_U$ represents the mean over all unitary transformations in the group and 
\begin{equation}
  \rho^e = \Big\langle U^\dagger \rho \ U \Big\rangle_U\,.
  \label{eq:rhoe}
\end{equation}
is the effective density matrix given as the `singlet' projection of \(\rho\) under the unitary group. Stated in words, if we only have access to operators that are invariant under the unitary group, then \cref{eq:effective} means that we have access only to this projection of \(\rho\) and so we can replace it by \(\rho^e\) in all calculations. Note that we can choose the mean in Eq.~\eqref{eq:rhoe} to be over a group invariant measure, so that evolution of \(\rho\) by any unitary \(\hat V \in \{U\}\), such that $\rho \to V\ \rho V^\dagger$, will leave \(\rho^e\) unchanged since
\begin{align}
\langle U^\dagger \hat V\rho \hat V^\dagger U \rangle_U\ =\ \rho_e.
\end{align}
Because of this, projection of \(\rho\) to \(\rho^e\) commutes with any evolution of the form \(\prod_i \hat U_i\), where each \(\hat U_i\) is either an element of \(\{U\}\), or commutes with all of them. In our example, this means that \(\rho^e\) is not changed by evolutions that act independently on the subsystem of interest and its complement.

We can connect the concept of an effective density matrix to the standard idea of the reduced density matrix by considering the Hilbert space of the full system $S$ as a direct product of two subsystems $A$ and $C$. Let us write the operators and density matrices in a direct product basis. Thus \(\rho \equiv \sum_{aa',cc'}\rho_{ac;a'c'} (|a\rangle \otimes |c\rangle) (\langle a'| \otimes \rangle c'|)\). Then the standard definition of the reduced density matrix for the first subsystem given by
\begin{align}
\rho^r_A \ =\ \sum_{aa',c} \rho_{ac;a'c} |a\rangle \langle a'| 
\end{align}
On the other hand we can define the first subsystem as associated with observables of the form \( (\hat O \otimes \hat I) \), all of which are invariant under the unitary group \( (I \otimes U) \). It is then easy to see that\footnote{This can be easily shown in a single qubit system by parameterizing the $2\times 2$ 
unitary matrices as $u(\phi,\theta,\chi) \equiv \{(e^{i\phi} \cos\theta, e^{i\chi} \sin\theta),\allowbreak(-e^{-i\chi}\sin\theta,e^{-i\phi}\cos\theta)\}$. If we consider the Pauli basis of the \(2\times2\) matrices, then \((1/2\pi)^3  \int_0^{2\pi}d\phi\allowbreak \int_0^{2\pi}d\theta \allowbreak \int_0^{2\pi}d\chi\allowbreak\, u (I,\sigma^x,\sigma^y,\sigma^z) u^\dagger = (I,0,0,0)\), which then leads to the result.}
\begin{align}
\Big\langle\ (I \otimes U)^\dagger \ (|a\rangle \otimes |c\rangle)&(\langle a'| \otimes \rangle c'|)\  (I \otimes U)\  \Big\rangle_U
\nonumber \\
& =\ \delta_{cc'} \ (|a\rangle \otimes |c\rangle)(\langle a'| \otimes \rangle c'|)
\end{align}
From this we obtain
\begin{align}
\rho^e\ =\  \langle (I \otimes U)^\dagger \ \rho\ (I \otimes U)  \rangle_U \ =\ \rho^r_A \otimes I. 
\end{align}
The condition for the reduction to commute with the evolution, then, means that we need the evolution to be a product of evolutions acting on the two systems independently, {\it i.e.,} there is no interaction between the two subsystems.

%% file: app_perm.tex
\section{Permutation Symmetric Density Matrix}
\label{app:perm}

Consider the Hilbert space of the register of spins that Alice and Bob have. Let us label the basis states of this space as \(\bigotimes_{i=0,\ldots,N-1} |s_i\rangle\), where $s_i$ stands for $2b$ spins. Now, we study the permutation symmetric version of an \(n\)-subsystem operator on this space. For this we consider operators \(O_n\) that act nontrivially only on the first \(n\) subsystems. Let us define an operator
\begin{align}
O^{S,m}_{n} \equiv \Big\langle {\cal P}_m O_n {\cal P}^\dagger_m \Big\rangle_{{\cal P}_m}  
\end{align}
where \({\cal P}_m\) is an element of the group of permutation operators that permute the first $m \geq n$, $2b$ spins. Using the arguments of \cref{app:effdens} we can see that the operators \(O^{S,m}_n\) can only access to the reduced density matrix \(\rho^r_m\) where all subsystems beyond the \(m^{\rm th}\) have been traced over. Furthermore, because of the permutation symmetry, not even the whole reduced density matrix is accessible, but only the effective density matrix
\begin{align}
\rho^S_m = \Big\langle {\cal P}_m^\dagger \rho_m {\cal P}_m\Big\rangle_{{\cal P}_m}.
\end{align}
Given any set of operators \(\{O_n\}\), the same density matrices $\rho^S_m$ suffice to give not only the mean values of \(O_n\) over the permutation group, but also other permutation invariant functions such as their variances and all higher cumulants over the permutation group, and hence the entire joint distribution of any commuting subset of them.

The matrices \(\rho^S_m\) for \(n \leq m \leq N\) form a sequence of permutation symmetric density matrices such that \(\rho^S_m\) is the partial trace of \(\rho^S_{m+1}\) by construction. 
When $N$ is strictly infinite, by the quantum de Finetti theorem~\cite{hudson1976locally,QdF}, we can write \(\rho^S_m\) as an average over direct product of single system subsystems, {\it i.e.,}
\begin{equation}
  \rho^S_m \ =\ \int d\rho\  p(\rho) \ \bigotimes_m \rho
  \label{eq:definetti}
\end{equation}
for all \(m\), where \(p(\rho)\) is some probability density of density matrices $\rho$ of one subsystem, and \(\bigotimes_m \rho\) stands for the direct product of \(m\) copies of \(\rho\). As a final step, note that if the variance of every \(O_n\) over the permutation group of size \(m\) vanishes as \(m\to\infty\), as is required if the subsystems are generated by a process whose correlations are finite, then  \(p(\rho) = \delta(\rho-\bar{\rho})\) for some \(\bar{\rho}\).  In this case,  \(\bar{\rho}\) is the partial trace of $\rho^S_m$
over all but one subsystem, which can also be calculated as the average single subsystem reduced density matrix. Note that $\bar{\rho}$ does not depend on $m$ due to the properties of \(\rho^S_m\). 

\section{Reduced Density Matrix in the Broken Phase}
\label{sec:rdmbp}

The idea of a reduced density matrix in a broken phase of a physical system requires some thought since in this case the system develops long range correlations, which implies that the probability density $p(\rho)$ defined in \cref{eq:definetti} is no longer a single delta function and is a sum (or integral) over delta functions representing every ground state. As we discussed in \cref{app:effdens}, the notion of a reduced density matrix arises from the locality of operators that we wish to compute. In the thermodynamic limit, we implicitly consider quantities that are averaged over the whole system. This is why even if the operator we are interested in is defined only over a local region, the notion of a reduced density matrix over that region alone can become subtle.

To understand this further consider a system whose Hilbert space factorizes into an infinite number of identical local Hilbert spaces in the sense described in \cref{app:perm}. Consider now an operator \(O\) that acts only within each local Hilbert space and let \(O_R\) be the operator that acts on another identical local Hilbert space obtained by a translation $R$. Then we can define the concept of a mean operator $\langle O\rangle_R$ as an average over all $O$ over all the local Hilbert spaces. If this infinite system is described by a density matrix \(\rho\) then the quantum de Finneti theorem~\cite{hudson1976locally,QdF} says that  
\begin{align}
\Tr\rho\, (\langle O\rangle_R)^n \ =\ 
  \int d\rho' p(\rho') \Tr (\bigotimes_R \rho') (\langle O\rangle_R)^n.
\label{eq:defenniti-1}
\end{align}
Stated differently, since the left-hand side is invariant under the permuting the regions \(R\), we can use the arguments in \cref{app:perm} to replace \(\rho\) by \(\rho^S\) and use Eq.~\eqref{eq:definetti} to further write
\begin{equation}
  \rho^S = \int d\rho' p(\rho') \left(\bigotimes_R \rho'\right)\,,
  \label{eq:redprob}
\end{equation}
which allows us to obtain the $p(\rho')$ from $\rho$. Thus, \cref{eq:redprob} does allow us to calculate the ensemble distribution of any operator as long as the operator of interest is of of the form $\langle O\rangle_R$. When \(p(\rho') = \delta(\rho' - \rho^r)\) where $\rho^r$ is the reduced density matrix of each of the local region then we can use the usual notion of the reduced density matrix to compute our observables. 

Since we are interested in the negativity of a bipartite reduced density matrix obtained by tracing out the rest of the degrees of freedom on a thermodynamic system as discussed in \cref{sec:negativity}, we are assuming \(p(\rho') = \delta(\rho' - \rho^r)\) implicitly. This is correct in the symmetric phase where there is no long range order. On the other hand in the presence of long range order, cluster decomposition fails and \(p(\rho')\) is not a delta function that depends on a single reduced density matrix. So a more careful definition of negativity is necessary that expresses it entirely in terms of operators of the form
$(\langle O\rangle_R)^n$ in \cref{eq:defenniti-1}. Instead, we follow an alternate route: since the broken phase of the \ac{TFIM} describes two ground states we expect
\begin{align}
p(\rho') = \frac{1}{2}\Big(\delta(\rho' - \rho^r_+) + \delta(\rho' - \rho^r_-)\Big)
\end{align}
So it is natural to study the negativity using the idea of a reduced density matrices in each of the sectors. Furthermore, the two sectors are related by the Ising symmetry, and hence are expected to give the same results. So, we can choose to study any one of these sectors, and this can be achieved by pinning a boundary spin using a parallel field, thus breaking the degeneracy. In a finite volume, there is no exact symmetry breaking---the two ground states will mix and their energies will be different. The magnitude of the boundary field needs to be much larger than the ground-state splitting, but much less than the energy of the first excitation above them.  Since as the volume increases, the energy of ground-states splitting approaches zero exponentially, {\it i.e.,} much faster than the lowest excitation energy, the boundary spin can be made arbitrarily small in the limit. This is analogous to the standard technique of taking the infinite volume limit with a fixed parallel field, which is then taken to zero; instead of trying to take the infinite volume limit at zero field.  With degeneracy thus removed, and cluster decomposition restored, we can define \(\rho^{\rm red}\) as usual by partial trace.

%% file: app_mps.tex
\section{Matrix Product States}
\label{app:mps}

\begin{figure*}
  \centering
  \includegraphics[width=0.9\linewidth]{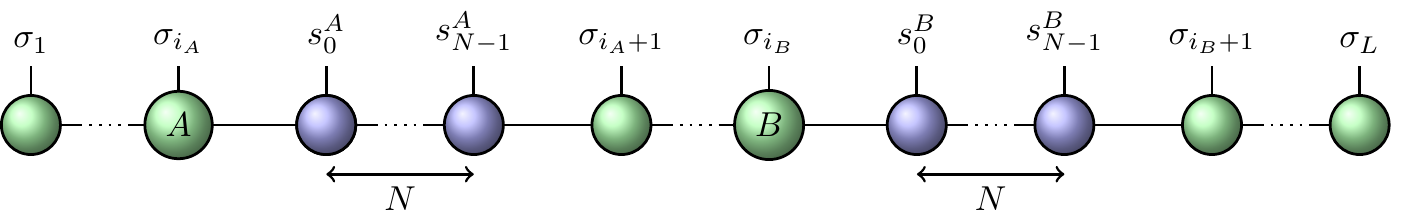}
\caption{The \ac{MPS} ansatz for our $(L+2N)$-spin system which include spins in the chain labeled as $\sigma_i,i=1,2...,L$ and $N$ spins in Alice's register $R_A$ (labeled $s^A_i,i=0,..,N-1$) and $N$ spins in the Bob's register $R_B$ (labeled $s^B_i,i=0,..,N-1$). Alice's register is placed next to the Alice's spin $\sigma_{i_A}$, and  Bob's register is placed next to Bob's spin $\sigma_{i_B}$. We choose to place the registers close to the respective spins to minimize the spread of entanglement when we perform the swaps. The spins in the spin chain are shown in green and the registers are shown in blue. Each node is a tensor, and each leg represents an index for the tensor. Legs which connect two tensors together represent indices which are summed over.}
  \label{fig:mps-structure}
\end{figure*}

We perform calculations using the variational approximation that the ground state of our system can be represented as a matrix product state (\ac{MPS}). For an $L$-site system, an \ac{MPS} is an ansatz for the many-body wavefunction of the form
\begin{align}
  |\psi\> = \sum_{\sigma_1, \dotsc, \sigma_{L}} \Tr M_1^{\sigma_1} M_2^{\sigma_2} \cdots M_{L}^{\sigma_L} | \sigma_1 \sigma_2 \cdots \sigma_{L} \>
\end{align}
where the $\sigma_i$ label an orthogonal basis for the local Hilbert space at site $i$, and each $M_i^{\sigma_n}$ is a matrix of dimensions $m_{i}\times m_{i+1}$. For arbitrary bond dimensions $m_{i}$, any state can be written in this form. However, for ground states of local gapped one-dimensional systems, it has been rigorously shown that \ac{MPS} provides an efficient approximation with a small bond dimensions \cite{schollwock_densitymatrix_2011}. This fact, combined with efficient algorithms such as \ac{DMRG} \cite{white_density_1992a,white_densitymatrix_1993}, which perform a variational search for the ground state within the \ac{MPS} space, gives a powerful suite of tools to study one-dimensional gapped systems. The algorithm optimizes the bond dimensions $m_i$ during the search. All our \ac{MPS} computations were performed using ITensor \cite{itensor,itensor-r0.3}.

When we study entanglement harvesting we introduce the Alice's register $R^A$ and Bob's register $R^B$ into the spin chain. Since we choose to work with $b=1$, our system consists not only of the spin-chain with $L$ sites but also two registers $R^A$ and $R^B$ each consisting of $N$ spins. To make an \ac{MPS} ansatz for the wavefunction of this combined system, we need to place all the $L+2N$ spins that belong to the spin-chain and the two registers in a single chain. To make sure that the interactions stay as local as possible, we insert Alice's register $R_A$ to the right of Alice's spin located at $i_A$ in the chain, and Bob's register $R_B$ to the right of Bob's spin located at $i_B$ as shown in \cref{fig:mps-structure}, the two spins located at $i_A$ and $i_B$ being separated by $\delta$. The insertions of registers naturally extends the $L$-site \ac{MPS} to an $(L+2N)$-site \ac{MPS}
\begin{align}
  \cdots M^{\sgA}_{i_A} \cdots &M^{\sgB}_{i_B} \cdots \\
                    &\mapsto \cdots M^{\sgA}_{i_A} \prod_{j=0}^{N-1} A^{\sA_j}_j  \cdots M^{\sgB}_{i_B}\ \prod_{k=0}^{N-1} B^{\sB_k}_k \ \cdots \\
\end{align}
where the newly inserted tensors $A^{\sA_j}$ and $B^{\sB_k}$ have initial bond dimensions $m_A\times m_A$ or $m_B\times m_B$ where $m_A$ and $m_B$ are bond dimensions on the original link of the spin-chain joining sites $(i_A,i_A+1)$ and $(i_B,i_B+1)$ respectively and are initially set to 
\begin{align}
A^{\sgA_i}_{m, m'} &= \delta_{\sgA_i, \uparrow}   \delta_{m, m'},\qquad
B^{\sgB_i}_{m, m'} &= \delta_{\sgB_i, \uparrow}   \delta_{m, m'}.
\end{align}
For register size $N$, this implies that the interactions are always between spins at most $N+1$ sites apart. This ensures that, for small register sizes, the interactions are local and the \ac{MPS} ansatz is applicable as we time-evolve the system. The bond dimensions of the various tensors are allowed to change with time evolution.

Our computation proceeds as follows. We begin with the $L+2N$-site \ac{MPS} and use \ac{DMRG} to find the ground state of the \ac{TFIM}. Therefore, if the ground state of the $L$-site Ising chain is $|\Omega(g)\>$, then our initial state is given by
\begin{align}
  |\psi^-(0)\> = |\Omega(g) \>\ \ |\! \uparrow \>^{\otimes N}_{\text{A}}\  |\! \uparrow \>^{\otimes N}_{\text{B}}.
\end{align}
In our protocol both Alice and Bob exchange one spin from the spin-chain into their registers during every measurement. This means we apply a swap gate as a unitary time evolution operator defined in \cref{eq:swap} between $\sgA\leftrightarrow \sA_0$ and $\sgB\leftrightarrow \sB_0$ to create the state $|\psi^+(0)\> = U^A_0U^B_0 |\psi^-(0)\>$. Then we let the state  $|\psi^+(0)\>$ evolve under the Ising model Hamiltonian (which only acts on the $L$ sites in the spin chain, leaving the registers untouched), for a time interval $t_{\rm min}(g)$ as discussed in \cref{sec:singleswap} 
\begin{align}
|\psi^-(1)\>\ =\ e^{-i H t_{\rm min}(g)} |\psi^+(0) \>,
\end{align}
and then apply a swap gate defined in \cref{eq:swap} between $\sgA\leftrightarrow \sA_1$ and $\sgB\leftrightarrow \sB_1$ to create the state $|\psi^+(1)\> = U^A_1U^B_1 |\psi^-(1)\>$. We repeat this process and obtain the state $|\psi^+(\alpha)\>$ after applying the swap gate between $\sgA\leftrightarrow \sA_\alpha$ and $\sgB\leftrightarrow \sB_\alpha$ after $\alpha+1$ swaps, where the swaps are done at times $\alpha t_{\rm min}(g)$. The density matrix $\rho^R_\alpha$ is then computed using the relation
\begin{align}
\rho^R_\alpha\ =\ \mathrm{Tr}(|\psi^+(\alpha)\>\<\psi^+(\alpha)|)
\end{align}
where the trace is over the Hilbert space of the $L$ spins that belong to the spin-chain. We perform the time evolution using a second-order Trotterization scheme.

In the broken phase of the \ac{TFIM} (i.e., $g < 1.0$), the ground state is doubly degenerate when $L\rightarrow \infty$. However since all of our calculations are performed for a finite value of $L$, there is a unique ground state, which can be considered as a symmetric superposition of the two ferromagnetic ground states.  Thus, the $\ZZ_2$ order parameter $\< Z_i \>$ at the site $i$ on the spin-chain will be zero for the true finite volume ground state. However, we would like to obtain results with one of the states which smoothly goes to one of the ground states in the infinite volume limit. Such a state should have a non-zero $\< Z_i \>$ in the broken phase.

\begin{figure}[htbp]
  \centering
  \includegraphics[width=0.5\linewidth]{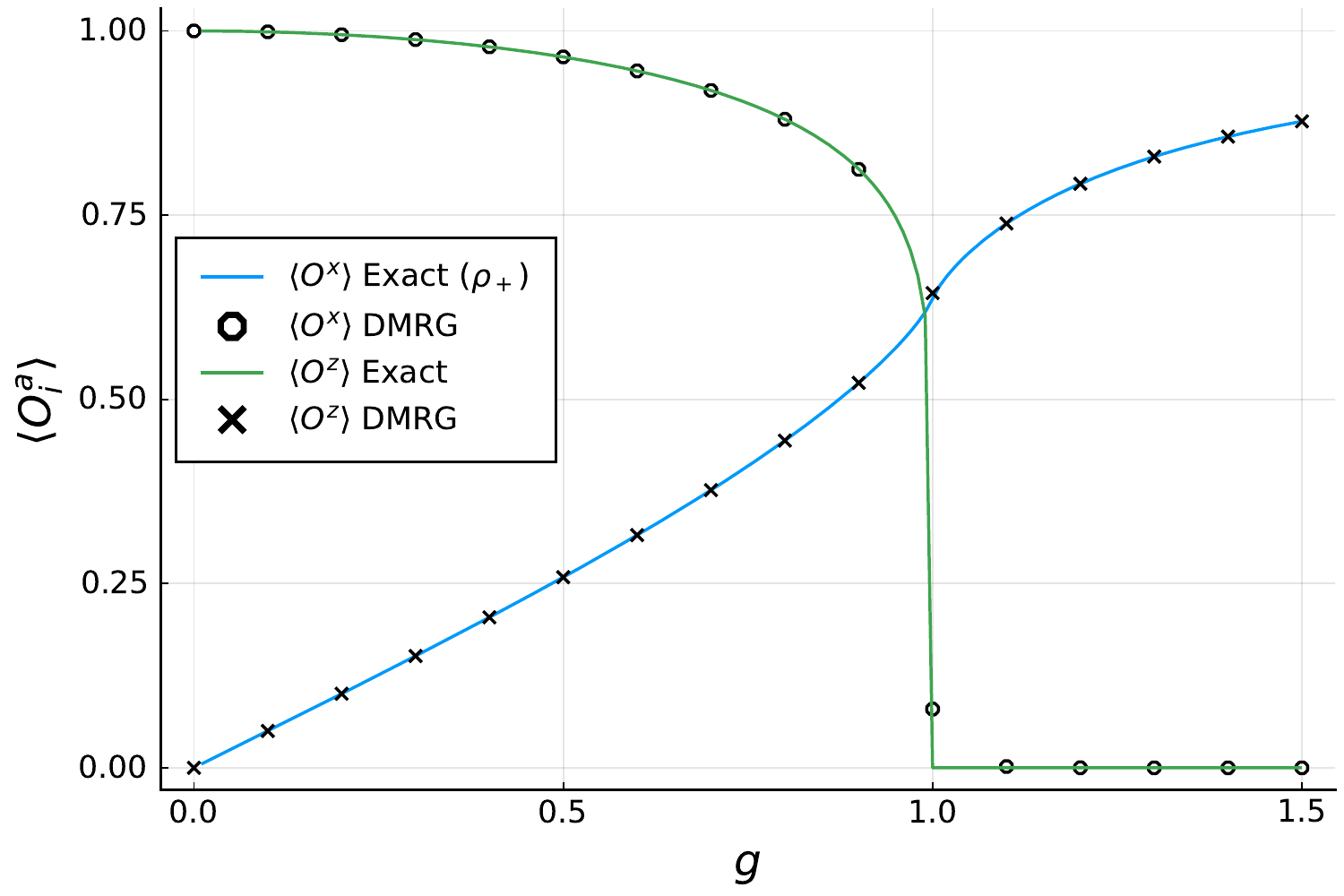}
  \caption{The expectation values $\langle O_i^z\rangle \equiv \langle Z_i\rangle$ and $\langle O_i^x\rangle \equiv \langle X_i\rangle$ as a function of $g$ where $i$ is chosen to the site at the center of the lattice.  The exact result using the density matrix $\rho$, as discussed in \cref{app:exact}, is shown as a solid line. The markers show the results obtained using \ac{DMRG} with a pinning field on a lattice with $L=64$.}
  \label{fig:1pt-corr}
\end{figure}

Since \ac{DMRG} is a variational optimization algorithm, we can force it to find one of the ferromagnetic ground state by using a small pinning field at the boundary, and additionally biasing the initial state to be close to, say, all up spins. Therefore, for \ac{DMRG}, we use the Hamiltonian
\newcommand\hpin{h_{\text{pin}}}
\begin{align}
 H = -\sum_{i=1}^{L-1} Z_{i}Z_{i+1} - g \sum_{i=1}^{L} X_i - \hpin(Z_1  + Z_{L}),
\end{align}
and set $\hpin = 0.01$. We find this to be sufficient to obtain the correct state for the values of $g, L$ considered in this work. For example, in \cref{fig:1pt-corr}, we show the expectation values $\< Z_i \>$ obtained using the pinning field for $L=64$, as well as the exact results (\cref{app:exact}) obtained in the thermodynamic limit as the pinning field is removed. The order parameter $\< Z_i \>$ shows the expected behavior as we tune $g$ through the phase transition, dropping sharply to zero at the phase transition. For $L=16$ considered in the calculations of the protocol, we find similar behaviour, although there are strong finite volume effects near the critical point ($g \gtrsim 0.75$) in the broken phase.

%% file: app_exact.tex
\section{Exact Results in the \texorpdfstring{\ac{TFIM}}{TFIM}}
\label{app:exact}

\begin{figure*}[htb]
  \centering
  \includegraphics[width=0.49\linewidth]{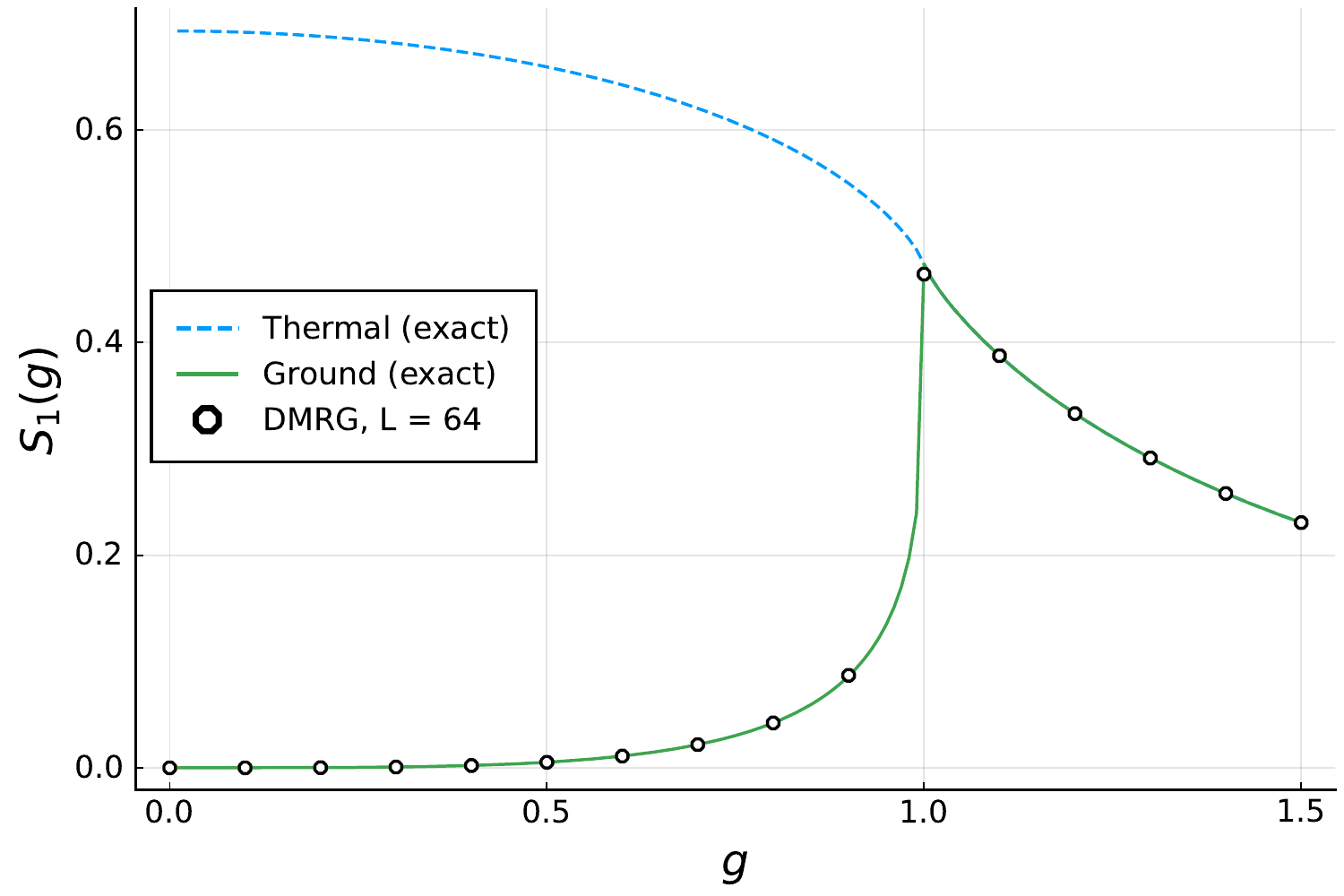}
  \includegraphics[width=0.49\linewidth]{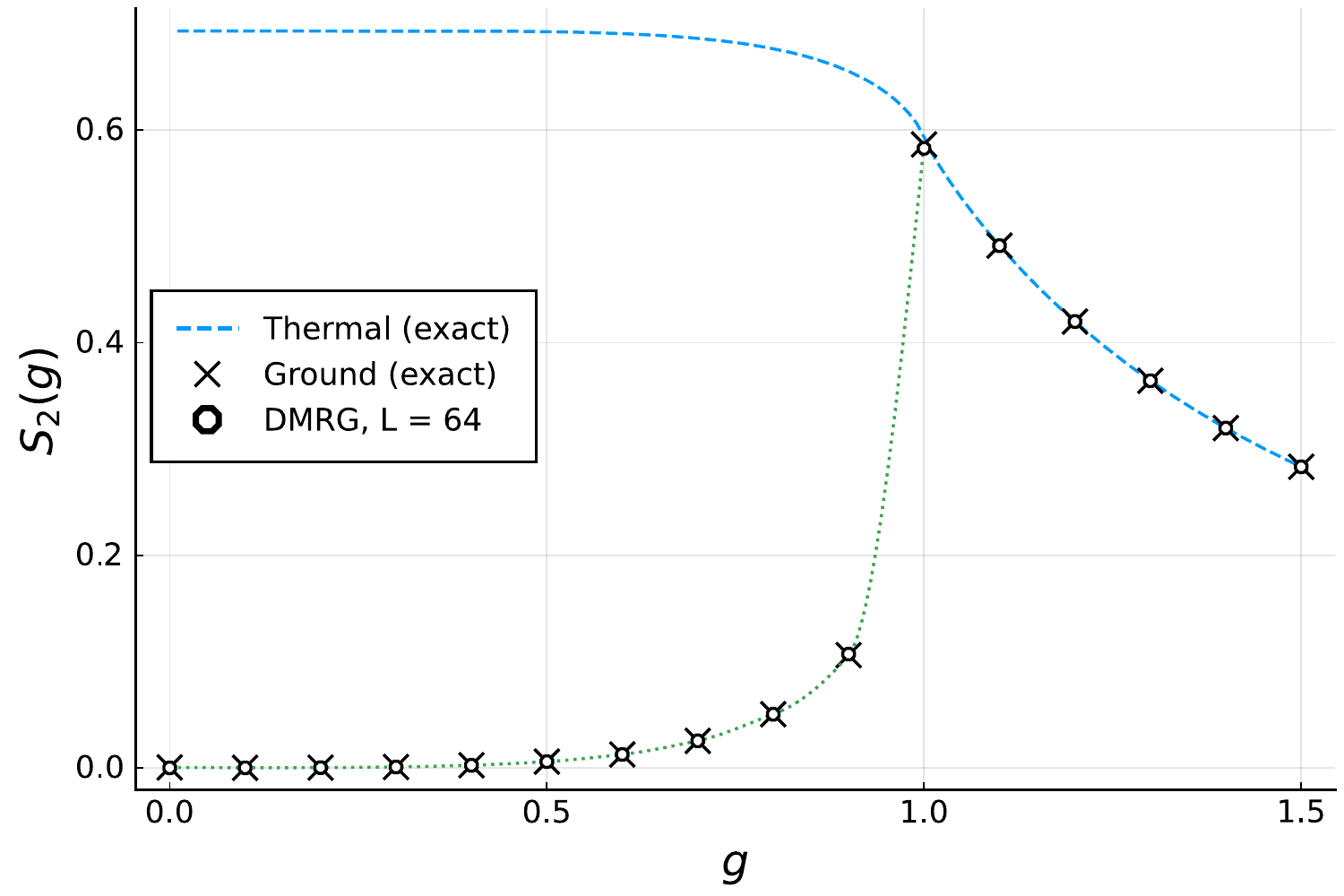}
  \caption{On the left panel, we show the single-site von Neumann entropy as a function of the coupling $g$ computed using the reduced density matrix $\rho^r_i$ of the ground state of the \ac{TFIM}, where $i$ is chosen to the site at the center of the lattice. As explained in the text, $\rho^r_i$ can be computed exactly using $\rho$ (solid line) or $\rho_{\rm th}$ (dashed line). The calculations using \ac{MPS} with $L=64$ are also shown as circles. On right panel, we show the two-site entanglement entropy as a function of $g$ computed using the reduced density matrix $\rho^r_{ij}$ of the ground state of the \ac{TFIM}, where $i,j$ are nearest neighbor sites at the center of the lattice. Again $\rho^r_{ij}$ can be computed using $\rho$ or $\rho_{\rm th}$. The exact answer obtained using $\rho_{\rm th}$ is known and is shown as the dashed line. The exact answer is not known for the ground state with a pinning field in the broken phase, but our answer obtained using \ac{MPS} is shown as circles.
  }
  \label{fig:entS}
\end{figure*}

In this section we wish to point out some subtleties in the calculation in the broken phase. First note that in our calculations we have assumed that the ground state of the Ising model is unique for all values of $g$ and $\rho$ is the density matrix for this unique ground state.  However, as discussed in \cref{sec:groundstate} and \cref{app:mps}, in the broken phase the ground state is doubly degenerate and we have to make a choice. We do this by introducing a small pinning field, taking the thermodynamic limit and then removing the pinning field. Thus, in our calculations $\rho$ corresponds to a single ground state density matrix for all values of $g$. In contrast one may in principle compute $\rho_{\rm th}$ to denote the thermal density matrix at zero temperature, which is different from $\rho$ in the broken phase since it is a density matrix of a mixed state with equal probabilities of being in both ground states. We want to be sure that the results we obtain using our \ac{MPS} approach in a finite volume with a small pinning field is from $\rho$ rather than $\rho_{\rm th}$. To confirm this, in this appendix we show our results for the von-Neumann entropy obtained from one-site and two-site reduced density matrices $\rho^r_i$ and $\rho^r_{ij}$ which are different depending on whether we use the $\rho$ or $\rho_{\rm th}$ to compute them and compare these results with exact results our \ac{DMRG} calculations.
For the \ac{TFIM} defined in \cref{eq:H-tfim} the expectation values
\begin{align}
C^a =   \mathrm{Tr}(\rho O^a_i),\ \ c^a =   \mathrm{Tr}(\rho_{\rm th} O^a_i),
\label{eq:op}
\end{align}
where $O^x_i = X_i$, $O^y_i = Y_i$ and $O^z_i = Z_i$, can be computed exactly. One can show that 
\begin{align}
  C^x &= \frac1\pi \int_0^\pi\!\!\! d \phi\ 
  \frac{(g + \cos \phi)}{(\sin\phi)^2 + (g + \cos \phi)^2}, \\
  C^z  &= \begin{cases}
        0 & g \geq 1, \\
        (1-g^2)^{1/8} & g < 1,
        \end{cases}
\end{align}
whereas $c^x=C^x$ and $c^z = 0$. Thus, we see that $C^z$ is different in the broken phase. This difference affects the single-site density matrix, which is given by either
\begin{align}
\rho^r_i &= \frac{1}{2} (\Id + C^x \sigma^x + C^z \sigma^z)
\end{align}
or a similar expression where the $C$'s are replaced by $c$'s depending on whether we use $\rho$ or $\rho_{\rm th}$ in computing the coefficients. In the left plot of \cref{fig:entS}, we show how the single site von-Neumann entropy obtained using $S(g) = Tr(\rho_i^r\ln(\rho^r_i))$ differs from $\rho$ and $\rho_{\rm th}$. The difference is only present in the broken phase $g<1$. Our \ac{DMRG} results agree with those calculated using $\rho$.

In the right plot of \cref{fig:entS}, we plot the two-site von Neumann entropy obtained from the two site reduced density matrix $\rho^r_{ij}$ where $|i-j|=1$ and $i,j$ are chosen at the center of the lattice to minimize finite size effects. We again notice that the difference between calculating $\rho^r_{ij}$ using $\rho$ or $\rho_{\rm th}$ in the broken phase ($g<1$) and our \ac{DMRG} results agree with those calculated using $\rho$. The exact results shown were computed using the formula known in the literature \cite{osborne_entanglement_2002}, where
\begin{align}
\rho^r_{ij} &= \frac{1}{4} \textstyle\sum_{\mu,\nu}\ C^{\mu\nu}_{ij} \sigma^\mu \otimes \sigma^\nu.
\label{eq:2srdm}
\end{align}
where the two point correlation functions
\begin{align}
 C^{ab}_{ij} =   \mathrm{Tr}(\rho O^a_i O^b_j),
\end{align}
have been computed using the density matrix $\rho$. We use a similar expression to compute the thermal reduced density matrix, but $C^{\mu\nu}_{ij}$ are replaced by $c^{\mu\nu}_{ij}$ computed using $\rho_{\rm th}$ instead of $\rho$. In \cref{eq:2srdm} we allow $\mu,\nu=0,\dotsc,3$ and $\sigma^0 = \Id$ and $\sigma^i,i=1,2,3$ are the Pauli matrices. Note that the correlation functions $C^{\mu\nu}_{ij}$ and $c^{\mu\nu}_{ij}$ depend only on $|i-j|$ and are symmetric under the exchange $a\leftrightarrow b$. The diagonal ones can be evaluated exactly in terms of the function $G(r)$ given by
\begin{align}
  G(r) = \frac{1}{\pi}
  \int_0^{\pi}\!\!\!\! d \phi\  
  \frac{(\cos(\phi r)(g + \cos \phi) - \sin(\phi r)\sin\phi}{\sqrt{(\sin\phi)^2 + (g + \cos \phi)^2}}.
\end{align}
In particular we can show
\begin{align}
  C^{xx}_{ij} &= C_i^xC_j^x - G(i-j)G(j-i) \\
  C^{yy}_{ij} &= \det \GGt^{[r]} \\
  C^{zz}_{ij} &= \det \GG^{[r]}.
\end{align}
where we have defined $r \times r$ matrices $\GG^{[r]}$ and $\GGt^{[r]}$ as
\begin{align}
  \GG_{a,b}^{[r]} = G(a - b -1 ) \\
  \GGt_{a,b}^{[r]} = G(a - b + 1 ).
\end{align}
In the notation above, $a,b=1,\dotsc,r$ and $r = |i-j|$ represents the distance between the sites. Note further that $C^{00}_{ij}=1$, $C^{0a}_{ij}=C^a_i = C^a_j$ and $C^{xy}_{ij} = C^{yz}_{ij} = 0$. The mixed correlation function $C^{xz}_{ij} \neq 0$ in the broken phase and unfortunately it is not easy to obtain this function analytically. Further $C^{ab}_{ij} = c^{ab}_{ij}$ in all cases except for the fact that $c^{xz}_{ij}=0$ for all values of $g$ unlike $C^{xz}_{ij}$ . Therefore,  calculations for the reduced density matrix $\rho^r_{ij}$, will differ between the thermal ensemble and the fixed ground state with a pinning field. In our calculations we compute $C^{xz}_{ij}$ using \ac{DMRG} and plug it into the expressions in \cref{eq:2srdm}, but use analytical expressions for all other correlation functions.

In \cref{fig:neg-transition} we showed our results for the two-site entanglement negativity ${\cal N}(\rho^r_{ij})$ (with $b=1$) that we obtained using \ac{DMRG} and compared it with the exact results computed in Ref.~\cite{osborne_entanglement_2002}. We also observed that the negativity did not depend on whether we used $\rho$ or $\rho_{\rm th}$ in our calculations. We later discovered the reason for this is that the partial transpose of $\rho^r_{ij}$ required during the calculation of the ${\cal N}(\rho^r_{ij})$  simply corresponds to setting $C^{yy}_{ij} \mapsto -C^{yy}_{ij}$ (or $c^{yy}_{ij} \mapsto -c^{yy}_{ij}$. This leads to only one negative eigenvalue of the partial transpose which is also the negativity, and is given by
\begin{align}
  {\cal N}(\rho^r_{ij}) = \frac14 (1 - C^{xx}_{ij} + C^{yy}_{ij} - C^{zz}_{ij}).
\end{align}
Since this is independent of $C^{z}$ as well as $C^{xz}_{ij}$, the negativity is the same for both the thermal and the single ground state density matrices.  In other words, the two-point negativity is the same for both $\rhog$ and $\rhoth$, even though the von Neumann entropy is different in the broken phase.


%% file: app_alignment.tex
\section{Aligning Density Matrices}
\label{app:align}

A maximally entangled 2-qubit state can be constructed as
\begin{align}
\ket{\theta,\phi,\beta}=(R^\dagger_{\theta,\phi,\beta}\otimes R_{\theta,\phi,\beta}) (\ket{\uparrow}\otimes\ket{\downarrow}-\ket{\downarrow}\otimes\ket{\uparrow})\,,
\label{eq:BellParameters}
\end{align}
where the \(SU(2)\) rotation matrix is parameterized as
\begin{align}
R_{\theta,\phi,\beta} &= \begin{pmatrix}e^{i\phi}\cos \theta&-ie^{i\beta}\sin\theta\\
                                 -ie^{-i\beta}\sin\theta&e^{-i\phi}\cos\theta
                                 \end{pmatrix}\,,
\end{align}
Hence, we see that two density matrices can be maximally entangled but their average can be less entangled because these density matrices are aligned differently. This can lead to a loss of entanglement in the symmetrization step we discussed earlier in \cref{app:perm}. A way to check how close a density matrix $\rho$ is to a maximally entangled state is to measure the maximal singlet fraction \(F_m(\rho) =\bra{\theta,\phi,\beta}\rho\ket{\theta,\phi,\beta}\) \cite{Horodecki:1998zh}, which is nothing but the maximal fidelity of the given density matrix to any maximally entangled state \citet{Bennett:1996gf}.

Note that since the diagonal subgroup of \(U(2)\otimes U(2)\) leaves the singlet invariant, we can use \(I\otimes R_{\theta,\phi,\beta}^2\) instead of 
\(R_{\theta,\phi,\beta}^\dagger\otimes R_{\theta,\phi,\beta}\). In addition to the periodicities in \(\theta\to\theta+\pi\), \(\phi\to\phi+2\pi\), and \(\beta\to\beta+2\pi\), the rotation matrix \(R_{\theta,\phi,\beta}^2\) has the following properties:
\begin{align}
R_{\theta,\phi,\beta}^2 &= - R_{\theta+\pi/2,\phi,\beta}^2 \nonumber\\
&= R_{-\theta,\phi,\beta+\pi}^2 \nonumber\\
&= R_{\theta,\phi+\pi,\beta+\pi}^2 \,.
\end{align}
Since the phase of the state is irrelevant in this calculation, the region \(\theta\in[0,\pi/4\mathclose[\), \(\phi\in\mathopen]-\pi/2,\pi/2]\), \(\beta\in\mathopen]-\pi,\pi]\) contains all the independent maximally entangled density matrices\footnotepunct{There are singularities at special points in this domain: thus, at \(\theta=0\), \(\beta\) is indeterminate; and at \(\theta=\pi/4,\ \phi=0\), the period in \(\beta\) becomes \(\pi\).}. Thus, the fidelity can be numerically maximized within this domain.

Given a density matrix $\rho$ and the rotation angles $\theta,\phi,\beta$ that maximises the singlet fraction $F_m(\rho)$, we can construct a new aligned density matrix $\rho^A = U^\dagger_A \rho U_A$ such that the rotation angles $\theta=0,\phi=0,\beta=0$ maximizes its singlet fraction. These aligned density matrices can then be used while symmetrizing the density matrix.